\documentclass[iop,appendixfloats]{emulateapj}

\shorttitle{LG monitoring. I}
\shortauthors{Saremi et al.}
\usepackage[usenames,dvipsnames]{color}
\usepackage{amsmath}
\usepackage{float}
\usepackage{ctable} 
\usepackage{graphicx}
\usepackage{natbib}
\begin{document}
                                
\title{The Isaac Newton Telescope monitoring survey of Local Group dwarf galaxies. I. Survey overview and first results for Andromeda I}

\author{Elham Saremi\altaffilmark{1},
Atefeh Javadi\altaffilmark{1}, 
Jacco~Th.~van~Loon\altaffilmark{2},
Habib Khosroshahi\altaffilmark{1},
Alireza Molaeinezhad\altaffilmark{3,4},
Iain McDonald\altaffilmark{5},
Mojtaba Raouf\altaffilmark{6},
Arash Danesh\altaffilmark{7},
James~R.~Bamber\altaffilmark{2},
Philip Short\altaffilmark{8},
Lucia Su\'arez-Andr\'es\altaffilmark{9},
Rosa Clavero\altaffilmark{3},
Ghassem Goz­aliasl\altaffilmark{10,11,12}}

\altaffiltext{1}{School of Astronomy, Institute for Research in Fundamental Sciences (IPM), P.O.~Box 1956836613, Tehran, Iran; saremi@ipm.ir}
\altaffiltext{2}{Lennard-Jones Laboratories, Keele University, ST5 5BG, UK}
\altaffiltext{3}{Instituto de Astrof\'{\i}sica de Canarias, Calle V\'{\i}a L\'actea s/n, E-38205 La Laguna, Tenerife, Spain}  
\altaffiltext{4}{Departamento de Astrof\'{\i}sica, Universidad de La Laguna, E-38200 La Laguna, Tenerife, Spain}
\altaffiltext{5}{Jodrell Bank Centre for Astrophysics, Alan Turing Building, University of Manchester, M13 9PL, UK}
\altaffiltext{6}{Korea Astronomy and Space Science Institute, 776 Daedeokdae-ro, Yuseong-gu, Daejeon 34055, Republic of Korea}
\altaffiltext{7}{Iranian National Observatory, Institute for Research in Fundamental Sciences (IPM), Tehran, Iran}
\altaffiltext{8}{Institute for Astronomy, University of Edinburgh, Royal Observatory, Blackford Hill, Edinburgh EH9 3HJ, UK}
\altaffiltext{9}{Isaac Newton Group, Apartado de correos 321, E-38700 Santa Cruz de La Palma, Canary Islands, Spain}
\altaffiltext{10}{Finnish Centre for Astronomy with ESO (FINCA), Quantum,University of Turku, Vesilinnantie 5, 20014 Turku, Finland}
\altaffiltext{11}{Department of Physics, University of Helsinki, P.O.~Box 64, 00014 Helsinki, Finland}
\altaffiltext{12}{Helsinki Institute of Physics, University of Helsinki, P.O.~Box 64, 00014 Helsinki, Finland}
%@@@@@@@@@@@@@@@@@@@@@@@@@@@@@@@@@@@@@@@@@@@@@@@@@@@@@@@@@@@@@@@@@@@@@ abstract
%
\begin{abstract}

An optical monitoring survey in the nearby dwarf galaxies was carried out with the 2.5-m Isaac Newton Telescope (INT). 55 dwarf galaxies and four isolated globular clusters in the Local Group (LG) were observed with the Wide Field Camera (WFC). The main aims of this survey are to identify the most evolved asymptotic giant branch (AGB) stars and red supergiants at the end-point of their evolution based on their pulsational instability, use their distribution over luminosity to reconstruct the star formation history, quantify the dust production and mass loss from modelling the multi-wavelength spectral energy distributions, and relate this to luminosity and radius variations. In this first of a series of papers, we present the methodology of the variability survey and describe the photometric catalogue of Andromeda\,I (And\,I) dwarf galaxy as an example of the survey, and discuss the identified long period variable (LPV) stars. We detected 5\,581 stars and identified 59 LPV candidates within two half-light radii of the centre of And\,I. The amplitudes of these candidates range from 0.2 to 3 mag in the $i$-band. 75\% of detected sources and 98\% of LPV candidates are detected at mid-infrared wavelengths. We show evidence for the presence of dust-producing AGB stars in this galaxy including five extreme AGB (x-AGB) stars, and model some of their spectral energy distributions. A distance modulus of $24.41$ mag for And\,I was determined based on the tip of the red giant branch (RGB). Also, a half-light radius of $3.2\pm0.3$ arcmin is calculated.

\end{abstract}
%%%%%%%%%%%%%%%%%%%%%%%%%%%%%%%%%%%%%%%%%%%%%%%%%%%%%%%%%%%%%%%%%%%%%%
\keywords{
stars: evolution --
stars: AGB and LPV--
stars: luminosity function, mass function --
stars: mass-loss --
stars: oscillations --
galaxies: individual: And\,I --
galaxies: stellar content --
galaxies: Local Group}

%@@@@@@@@@@@@@@@@@@@@@@@@@@@@@@@@@@@@@@@@@@@@@@@@@@@@@@@@@@@@@@@@@@@@ section 1
%
\section{Introduction}

Presenting a complete suite of galactic environments, dwarf galaxies are the most abundant type of galaxies in the Universe. Due to their proximity, variety, and a wide range of metallicity (0.002 Z$_\odot$ to 0.08 Z$_\odot$; Boyer et al.\ 2015a), the Local Group (LG) dwarf galaxies offer a unique opportunity to study the connection between stellar populations and galaxy evolution. On one hand, with so many dwarf galaxies known in the LG, it has become possible to conduct comparative studies of different morphological types including dwarf spheroidals (dSphs), dwarf irregulars (dIrrs), and transition (dIrr/dSph, or dTrans) galaxies (e.g., Dolphin et al.\ 2005; Read et al.\ 2006; Tolstoy et al.\ 2009; Boyer et al.\ 2015a). On the other hand, recent studies on some of the dwarf galaxies in the local Universe show that there is an overlap in the structural properties, size--luminosity, and the surface brightness--luminosity relations between the dwarfs in and around the LG (e.g., Dalcanton et al.\ 2009; Weisz et al.\ 2011; McConnachie 2012; Karachentsev et al.\ 2013). So, the LG dwarfs can be used as representatives for the population of dwarf galaxies in the Universe at large (Mayer 2011).

Although, in recent years, the dwarf galaxies of the LG have been the focus of intense studies (e.g., Tolstoy et al.\ 2009; McConnachie 2012; Weisz et al.\ 2014; Boyer et al.\ 2015a), many open questions about galaxy evolution remain, such as to what extent and when gas removal mechanisms quenched their star formation. Can dwarf galaxies be rejuvenated? Can gas be (re-)accreted into dwarf galaxies? To what extent is stellar death able to replenish the interstellar medium (ISM) with metals and dust, and to what extent does it heat it and drive galactic winds?

The star formation history (SFH) is one of the most powerful tracers of galaxy formation and evolution and can help answer the above questions. This is because stars are testimony of the ISM to congregate and collapse within dark matter halos; they are also the main actors within galaxies as they drive much of the feedback upon the ISM, thus regulating further evolution. In the final stages of evolution, stars with birth masses ranging from 0.8 to 8 M$_\odot$ ascend the asymptotic giant branch (AGB) where they burn hydrogen and helium in concentric shells around a degenerate C/O core (Habing \& Olofsson 2003). After a short period of time, the helium shell becomes unstable and the star will undergo a Thermal Pulse (TP; Marigo et al.\ 2013). During this phase, TP-AGB, the convective envelope of the star can penetrate into deeper layers, efficiently bringing the products of nuclear burning to the stellar surface (third dredge-up). Third dredge-up increases the C/O abundance ratio of the envelope and might exceed unity resulting in a carbon star (Iben 1975; Sugimoto \& Nomoto 1975; Groenewegen \& de Jong 1993).

Stars with birth masses of $\sim$5--10 M$_\odot$ become the brighter super-AGB stars ($M_{\rm bol} \lesssim -7$ mag; Siess 2007). These stars experience Hot Bottom Burning (HBB; Iben \& Renzini 1983) at the base of the convective envelope. Here, dredged-up carbon is burnt into nitrogen and oxygen, hence these stars remain oxygen-rich. Red supergiants (RSGs) are brighter still, and come from even more-massive progenitors, up to $\sim$30 M$_\odot$, ending their lives as supernovae (Levesque et al.\ 2005; Levesque 2010; Yang \& Jiang 2012).

Instabilities between gravitation and radiation pressure (Lattanzio \& Wood 2004) at the end of the TP-AGB phase cause long period variability (LPV). LPVs have radial pulsations in their atmospheric layers with periods of hundreds of days (e.g., Wood 1998; Whitelock et al.\ 2003; Yuan et al.\ 2018; Goldman et al.\ 2019). The pulsation drives shocks through the circumstellar envelope, dissipating in regions where density and temperature are suitable for dust grain formation. Radiation pressure upon these grains then helps drive a stellar wind (Gehrz \& Woolf 1971; Whitelock et al.\ 2003; Goldman et al.\ 2017). These winds cause AGB stars to lose up to 80\% of their mass to the ISM (Bowen 1988; Bowen \& Willson 1991; Vassiliadis \& Wood 1993; van Loon et al.\ 1999, 2005; Cummings et al.\ 2018). Although RSGs contribute less to the total dust budget of a galaxy, their mass loss sets the conditions within which the ensuing supernova develops (van Loon 2010).

AGB stars and RSGs are unique objects for reconstructing the SFH of a galaxy, tracing stellar populations from as recently formed as 10 Myr ago to as ancient as 10 Gyr, because they are in the final stages of their evolution and their luminosity is directly related to their birth mass (Javadi et al.\ 2011a). Various efforts have been made to characterise AGB populations and RSGs in LG galaxies beyond the Magellanic Clouds at infrared (IR) wavelengths (Javadi et al.\ 2011a, 2014; Battinelli \& Demers 2013; Menzies, Feast \& Whitelock 2015; Boyer et al.\ 2009, 2015a, 2015b), however, only a small fraction of those stars were identified in optical surveys. The amplitude of these LPVs in optical bands are larger and thus easier to detect; especially for RSGs which tend to have small amplitudes anyway. Also, optical photometry better constraints the circumstellar dust envelope optical depth (and thus the dust production and mass-loss rate estimates). Finally, an optical survey gives us access to temperature variations and therefore radius variations of LPVs.

With this in mind, we conducted an optical survey of nearby galaxies (the most complete sample so far) with the 2.5-m Isaac Newton Telescope (INT) over nine epochs. Our main objectives include: identify all LPVs in the dwarf galaxies of the LG accessible in the Northern hemisphere, and then determine the SFHs from their luminosity distribution, using the method that we successfully applied to a number of LG galaxies previously (Javadi et al.\ 2011b, 2017; Rezaeikh et al.\ 2014; Hamedani Golshan et al.\ 2017; Hashemi et al.\ 2018; Saremi et al.\ 2019a); obtain accurate time-averaged photometry for all LPVs; obtain the pulsation amplitude of them; determine their radius variations; model their spectral energy distributions (SEDs); and study their mass loss as a function of stellar properties such as mass, luminosity, metallicity, and pulsation amplitude.

This first paper in the series presents the status of the monitoring survey of the LG dwarf galaxies along with first results for the Andromeda\,I (And\,I) dwarf galaxy (Saremi et al.\ 2019b) to introduce the methodology and scientific potential of this project. And\,I is a bright dSph ($M_{\rm V}= -11.7\pm0.1$ mag; McConnachie 2012) that was initially discovered on photographic plates by van den Bergh (1972). It lies some $3\rlap{.}^{\circ}3$ from the centre of M\,31 at a position angle of $\sim135^\circ$ relative to the M\,31 major axis (DaCosta et al.\ 1996). The distance to And\,I was determined {\it via} several methods: Based on the tip of the red giant branch (RGB), McConnachie et al.\ (2005) found a distance modulus of $\mu=24.36\pm0.05$ mag. Also, Conn et al.\ (2012) with similar method obtained $\mu=24.31\pm0.07$ mag. Mart\'{\i}nez-V\'azquez et al.\ (2017) estimated $\mu=24.49\pm0.12$ mag by using the RR\,Lyrae variables. Here we calculated a distance modulus of $24.41\pm0.05$ mag for And\,I based on the tip of the RGB.

This paper is organized as follows: In Sec.\ 2, we describe the survey and our observations. The data processing steps and data quality are described in Sec.\ 3. Our method to detect LPVs is presented in Sec.\ 4. In Sec.\ 5, we discuss the LPV candidates of And\,I and cross-correlate the data with other catalogues. A summary is given in Sec.\ 6.

%@@@@@@@@@@@@@@@@@@@@@@@@@@@@@@@@@@@@@@@@@@@@@@@@@@@@@@@@@@@@@@@@@@@@ section 2

%%%%%%%%%%%%%%%%%%%%%%%%%%%%%%%%%%%%%%%%%%%%%%%%%%%%%%%%%%%%%%%%%%%%%%%%%%%%%%%
% FIGURE 1
\begin{figure}
\includegraphics[width=1\columnwidth]{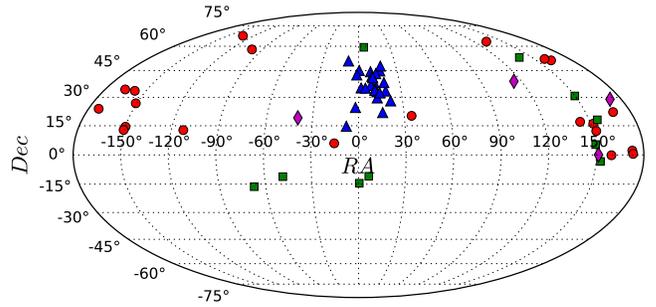}
\caption{The spatial distribution of the sample of galaxies. The red circles 
and blue triangles represent dwarf galaxies of the Milky\,Way and Andromeda, 
respectively. Isolated dwarfs are shown in green squares and GCs in purple 
lozenges. \label{fig:fig1}}
\end{figure}
%%%%%%%%%%%%%%%%%%%%%%%%%%%%%%%%%%%%%%%%%%%%%%%%%%%%%%%%%%%%%%%%%%%%%%%%%%%%%%%

%%%%%%%%%%%%%%%%%%%%%%%%%%%%%%%%%%%%%%%%%%%%%%%%%%%%%%%%%%%%%%%%%%%%%%%%%%%%%%%
% TABLE 1
\begin{table*}
\centering
\begin{minipage}{\textwidth}

\renewcommand{\thefootnote}{\alph{footnote}}
\end{minipage}

\begin{minipage}{\textwidth}
\caption{Observational properties of targets.}
\begin{tabular}{llllllllll}
\hline\hline
 \noalign{\smallskip}
{Galaxy}           &
{R.A.}             &
{Dec}              &
{$M_{\rm V}$}      & 
{$r_{\rm h}$\footnotemark[1]}    &
{Ellipticity\footnotemark[2]}    &
{$(m-M)_0$}        &
{$M_{\rm stars}$}  &
{[Fe/H]}           &
{type}             \\
&
{(J2000)}          &
{(J2000)}          &
{(mag)}            &
{(arcmin)}         &
&
{(mag)}            &
{$10^6M_\odot$}&
&
\\
 \noalign{\smallskip}
\hline
 \noalign{\smallskip}
\multicolumn{10}{l}{$\star\star$ Dwarf galaxies of Andromeda\dotfill}\\ 
And\,I     & 00 45 39.8 & $+38$ 02 28 & $-11.7\pm0.1$ & $3.20\pm0.30$ & $0.22\pm0.04$ & $24.41\pm0.05$ & 3.9   & $-1.45\pm0.04$  & dSph \\
And\,II    & 01 16 29.8 & $+33$ 25 09 & $-12.4\pm0.2$ & $6.20\pm0.20$ & $0.20\pm0.08$ & $24.07\pm0.06$ & 7.6   & $-1.64\pm0.04$  & dSph \\
And\,III   & 00 35 33.8 & $+36$ 29 52 & $-10.0\pm0.3$ & $2.20\pm0.20$ & $0.52\pm0.02$ & $24.37\pm0.07$ & 0.83  & $-1.78\pm0.04$  & dSph \\
And\,V     & 01 10 17.1 & $+47$ 37 41 & $-9.1\pm0.2$  & $1.40\pm0.20$ & $0.18\pm0.05$ & $24.44\pm0.08$ & 0.39  & $-1.6 \pm0.3 $  & dSph \\
And\,VI    & 23 51 46.3 & $+24$ 34 57 & $-11.3\pm0.2$ & $2.30\pm0.20$ & $0.41\pm0.03$ & $24.47\pm0.07$ & 2.8   & $-1.3 \pm0.14$  & dSph \\
And\,VII   & 23 26 31.7 & $+50$ 40 33 & $-12.6\pm0.3$ & $3.50\pm0.10$ & $0.13\pm0.04$ & $24.41\pm0.10$ & 9.5   & $-1.40\pm0.30$  & dSph \\
And\,IX    & 00 52 53.0 & $+43$ 11 45 & $-8.1\pm1.1$  & $2.50\pm0.10$ &  \dotfill     & $24.42\pm0.07$ & 0.15  & $-2.2 \pm0.2 $  & dSph \\
And\,X     & 01 06 33.7 & $+44$ 48 16 & $-7.6\pm1.0$  & $1.30\pm0.10$ & $0.44\pm0.06$ & $24.23\pm0.21$ & 0.096 & $-1.93\pm0.11$  & dSph \\
And\,XI    & 00 46 20.0 & $+33$ 48 05 & $-6.9\pm1.3$  & $0.71\pm0.03$ &  \dotfill     & $24.40^{+0.2}_{-0.5}$ &0.049&$-2.0\pm0.2$& dSph \\
And\,XII   & 00 47 27.0 & $+34$ 22 29 & $-6.4\pm1.2$  & $1.20\pm0.20$ &  \dotfill     & $24.70\pm0.30$ & 0.031 & $-2.1 \pm0.2 $  & dSph \\
And\,XIII  & 00 51 51.0 & $+33$ 00 16 & $-6.7\pm1.3$  & $0.78\pm0.08$ &  \dotfill     & $24.80^{+0.1}_{-0.4}$ &0.041&$-1.9\pm0.2$& dSph \\
And\,XIV   & 00 51 35.0 & $+29$ 41 49 & $-8.4\pm0.6$  & $1.70\pm0.80$ & $0.31\pm0.09$ & $24.33\pm0.33$ & 0.20  & $-2.26\pm0.05$  & dSph \\
And\,XV    & 01 14 18.7 & $+38$ 07 03 & $-9.4\pm0.4$  & $1.21\pm0.05$ & $0.00$        & $24.00\pm0.20$ & 0.49  & $-1.8 \pm0.2 $  & dSph \\
And\,XVI   & 00 59 29.8 & $+32$ 22 36 & $-9.2\pm0.4$  & $0.89\pm0.05$ & $0.00$        & $23.60\pm0.20$ & 0.41  & $-2.1 \pm0.2 $  & dSph \\
And\,XVII  & 00 37 07.0 & $+44$ 19 20 & $-8.7\pm0.4$  & $1.24\pm0.08$ & $0.27\pm0.06$ & $24.50\pm0.10$ & 0.26  & $-1.9 \pm0.2 $  & dSph \\
And\,XVIII & 00 02 14.5 & $+45$ 05 20 & $-9.7\pm0.1$  & $0.92\pm0.06$ &  \dotfill     & $25.66\pm0.13$ & 0.63  & $-1.8 \pm0.1 $  & dSph \\
And\,XIX   & 00 19 32.1 & $+35$ 02 37 & $-9.2\pm0.6$  & $6.20\pm0.10$ & $0.17\pm0.02$ & $24.85\pm0.13$ & 0.43  & $-1.9 \pm0.1 $  & dSph \\
And\,XX    & 00 07 30.7&$+35$ 07 56&$-6.3^{+1.1}_{-0.8}$&$0.53^{+0.14}_{-0.04}$&$0.30\pm0.15$&$24.52^{+0.74}_{-0.24}$&0.029&$-1.5\pm0.1$&dSph\\
And\,XXI   & 23 54 47.7 & $+42$ 28 15 & $-9.9\pm0.6$  & $3.50\pm0.30$ & $0.20\pm0.07$ & $24.67\pm0.13$ & 0.76  & $-1.8 \pm0.2 $  & dSph \\
And\,XXII  & 01 27 40.0 & $+28$ 05 25 & $-6.5\pm0.8$  & $0.94\pm0.10$ & $0.56\pm0.11$ & $24.82^{+0.07}_{-0.31}$ &0.034& $-1.8$   & dSph \\
M\,110     & 00 40 22.1 & $+41$ 41 07 & $-16.5\pm0.1$ & $2.46\pm0.1$  & $0.43\pm0.10$ & $24.58\pm0.07$ &  330  & $-0.8 \pm0.2 $  & dE   \\
M\,32      & 00 42 41.8 & $+40$ 51 55 & $-16.4\pm0.2$ & $0.47\pm0.05$ & $0.25\pm0.02$ & $24.53\pm0.21$ &  320  & $-0.25$         & dE   \\
Pisces\,I  & 01 03 55.0 & $+21$ 53 06 & $-10.1\pm0.1$ & $2.10\pm0.20$ & $0.20$        & $24.43\pm0.07$ & 0.96  & $-2.10\pm0.22$  &dTrans\\
Pegasus    & 23 28 36.3 & $+14$ 44 35 & $-12.2\pm0.2$ & $2.10$        & $0.46\pm0.02$ & $24.82\pm0.07$ & 6.6   & $-1.4\pm0.2$    &dTrans\\
&&&&&&&& \\
\multicolumn{10}{l}{$\star\star$  Dwarf galaxies of the Milky\,Way\dotfill}\\
Segue\,I   & 10 07 04.0 & $+16$ 04 55 & $-1.5\pm0.8$  & $4.4^{+1.2}_{-0.6}$&$0.48\pm0.13$ &$16.80\pm0.20$&0.00034&$-2.72\pm0.40$ & dSph \\
Segue\,II  & 02 19 16.0 & $+20$ 10 31 & $-2.5\pm0.3$  & $3.40\pm0.20$ & $0.15\pm0.10$ & $17.70\pm0.10$ &0.00086& $-2.00\pm0.25$  & dSph \\ 
Ursa\,Minor& 15 09 08.5 & $+67$ 13 21 & $-8.8\pm0.5$  & $8.20\pm1.20$ & $0.56\pm0.05$ & $19.40\pm0.10$ & 0.29  & $-2.13\pm0.01$  & dSph \\
UMa\,I     & 10 34 52.8 & $+51$ 55 12 & $-5.5\pm0.3$  & $11.30\pm1.70$& $0.80\pm0.04$ & $19.93\pm0.10$ & 0.014 & $-2.18\pm0.04$  & dSph \\
UMa\,II    & 08 51 30.0 & $+63$ 07 48 & $-4.2\pm0.6$  & $16.0\pm1.0$  & $0.63\pm0.05$ & $17.50\pm0.30$ & 0.0041& $-2.47\pm0.06$  & dSph \\
Willman\,1 & 10 49 21.0 & $+51$ 03 00 & $-2.7\pm0.8$  & $2.30\pm0.40$ & $0.47\pm0.08$ & $17.90\pm0.40$ & 0.0010& $-2.1$          & dSph \\
Coma\,Ber  & 12 26 59.0 & $+23$ 54 15 & $-4.1\pm0.5$  & $6.00\pm0.60$ & $0.38\pm0.14$ & $18.20\pm0.20$ & 0.0037& $-2.60\pm0.05$  & dSph \\
CVn\,I     & 13 28 03.5 & $+33$ 33 21 & $-8.6\pm0.2$  & $8.90\pm0.40$ & $0.39\pm0.03$ & $21.69\pm0.10$ & 0.23  & $-1.98\pm0.01$  & dSph \\
CVn\,II    & 12 57 10.0 & $+34$ 19 15 & $-4.9\pm0.5$  & $1.60\pm0.30$ & $0.52\pm0.11$ & $21.02\pm0.06$ & 0.0079& $-2.21\pm0.05$  & dSph \\
Bootes\,I  & 14 00 06.0 & $+14$ 30 00 & $-6.3\pm0.2$  & $12.6\pm1.0$  & $0.39\pm0.06$ & $19.11\pm0.08$ & 0.029 & $-2.55\pm0.11$  & dSph \\
Bootes\,II & 13 58 00.0 & $+12$ 51 00 & $-2.7\pm0.9$  & $4.20\pm1.40$ & $0.21\pm0.21$ & $18.10\pm0.06$ & 0.0010& $-1.79\pm0.05$  & dSph \\
Bootes\,III& 13 57 12.0 & $+26$ 48 00 & $-5.8\pm0.5$  &  \dotfill     & $0.50$        & $18.35\pm0.1$  & 0.017 & $-2.1 \pm0.2 $  & dSph?\\
Hercules   & 16 31 02.0 & $+12$ 47 30 & $-6.6\pm0.4$  & $8.60^{+1.8}_{-1.1}$ & $0.68\pm0.08$ & $20.60\pm0.20$&0.037 & $-2.41\pm0.04$& dSph \\
Draco      & 17 20 12.4 & $+57$ 54 55 & $-8.8\pm0.3$  & $10.00\pm0.30$& $0.31\pm0.02$ & $19.40\pm0.17$ & 0.29  & $-1.93\pm0.01$  & dSph \\ 
Leo\,I     & 10 08 28.1 & $+12$ 18 23 & $-12.0\pm0.3$ & $3.40\pm0.30$ & $0.21\pm0.03$ & $22.02\pm0.13$ &  5.5  & $-1.43\pm0.01$  & dSph \\
Leo\,II    & 11 13 28.8 & $+22$ 09 06 & $-9.8\pm0.3$  & $2.60\pm0.60$ & $0.13\pm0.05$ & $21.84\pm0.13$ & 0.74  & $-1.62\pm0.01$  & dSph \\
Leo\,V     & 11 31 09.6 & $+02$ 13 12 & $-5.2\pm0.4$  & $2.60\pm0.60$ & $0.50\pm0.15$ & $21.25\pm0.12$ & 0.011 & $-2.00\pm0.2 $  & dSph \\
Leo\,IV    & 11 32 57.0 & $-00$ 32 00 & $-5.8\pm0.4$  & $4.60\pm0.80$ & $0.49\pm0.11$ & $20.94\pm0.09$ & 0.019 & $-2.54\pm0.07$  & dSph \\
Leo\,T     & 09 34 53.4 & $+17$ 03 05 & $-8.0\pm0.5$  & $0.99\pm0.06$ & $0.00$        & $23.10\pm0.10$ & 0.14  & $-1.99\pm0.05$  &dTrans\\
Pisces\,II & 22 58 31.0 & $+05$ 57 09 & $-5.0$        & $1.10\pm0.10$ & $0.40\pm0.10$ & $21.31\pm0.18$ & 0.0086& $-1.9$          & dSph \\
Sextans    & 10 13 03.0 & $-01$ 36 54 & $-9.27\pm0.58$&   \dotfill    &   \dotfill    & $19.90\pm0.06$ &   40  & $-2.1\pm0.1$    & dSph \\
&&&&&&&& \\
\multicolumn{10}{l}{$\star\star$ Isolated dwarf galaxies\dotfill}\\
IC\,10     & 00 20 17.3 & $+59$ 18 14 & $-15.0\pm0.2$ & $2.65$        & $0.19\pm0.02$ & $24.50\pm0.12$ &  86   & $-1.28$         & dIrr \\
WLM        & 00 01 58.2 & $-15$ 27 39 & $-14.2\pm0.1$ & $7.78$        & $0.65\pm0.01$ & $24.85\pm0.08$ &  43   & $-1.27\pm0.04$  & dIrr \\
Sag\,DIG   & 19 29 59.0 & $-17$ 40 41 & $-11.5\pm0.3$ & $0.91\pm0.05$ & $0.50$        & $25.14\pm0.18$ &  3.5  & $-2.1\pm0.2$    & dIrr \\
Aquarius   & 20 46 51.8 & $-12$ 50 53 & $-10.6\pm0.1$ & $1.47\pm0.04$ & $0.50\pm0.10$ & $25.15\pm0.08$ &  1.6  & $-1.3\pm0.2$    &dTrans\\
UGC\,4879  & 09 16 02.2 & $+52$ 50 24 & $-12.5\pm0.2$ & $0.41\pm0.04$ & $0.44\pm0.04$ & $25.67\pm0.04$ &  8.3  & $-1.5\pm0.2$    &dTrans\\
Cetus      & 00 26 11.0 & $-11$ 02 40 & $-11.2\pm0.2$ & $3.20\pm0.10$ & $0.33\pm0.06$ & $24.39\pm0.07$ &  2.6  & $-1.9\pm0.10$   & dSph \\
Leo\,A     & 09 59 26.5 & $+30$ 44 47 & $-12.1\pm0.2$ &   2.15        & $0.40\pm0.03$ & $24.51\pm0.12$ &  6.0  & $-1.4 \pm0.2 $  & dIrr \\
Leo\,P     & 10 21 45.1 & $+18$ 05 17 & $-9.41^{+0.17}_{-0.5}$ &\dotfill & 0.52 &$26.19^{+0.17}_{-0.5}$&  0.57 & $-1.8\pm0.1$    &dTrans\\
Sextans\,A & 10 11 00.8 & $-04$ 41 34 & $-14.3\pm0.1$ & $2.47$        & $0.17\pm0.02$ & $25.78\pm0.08$ &  44   & $-1.85$         & dIrr \\
Sextans\,B & 10 00 00.1 & $+05$ 19 56 & $-14.5\pm0.2$ & $1.06\pm0.10$ & $0.31\pm0.03$ & $25.77\pm0.03$ &  52   & $-1.6$          & dIrr \\
&&&&&&&& \\
\multicolumn{10}{l}{$\star\star$ Globular Clusters\dotfill}\\
Segue\,III & 21 21 31.1 & $+19$ 07 03 & $-0.0\pm0.8$  & $0.47\pm0.13$ &  \dotfill     & $16.1\pm0.1$   &\dotfill& $-1.7^{+0.1}_{-0.3}$ & GC \\
Sextans\,C & 10 05 31.9 & $+00$ 04 18 &     $-5.69$   &      0.65     &  \dotfill     &    19.95       &\dotfill&    $-1.63$     & GC   \\
Palomar\,4 & 11 29 15.8 & $+28$ 58 23 &     $-6.01$   &      0.51     &  \dotfill     &    20.21       &\dotfill&    $-1.41$     & GC   \\
NGC\,2419  & 07 38 08.47& $+38$ 52 56.8&    $-9.42$   &      0.89     &     0.03      &    19.83       &\dotfill&    $-2.15$     & GC   \\
 \noalign{\smallskip}
\hline
\end{tabular}
\small{Notes: The physical properties of the sample galaxies are mostly taken from McConnachie (2012) except for Leo\,P (McQuinn et al.\ 2015), Sextans (Lee et al.\ 2003; {\L}okas 2009) and the GCs: Sextans\,C, Palomar\,4, NGC\,2419 (Harris 1996; 2010 edition) and Segue\,III (Boyer et al.\ 2015). The distance modulus and the half-light radius of And\,I are calculated in this study.}
\footnotetext[1]{The half-light radius ($r_{\rm h}$) is the distance along the semi-major axis that contains half the light of the galaxy.}
\footnotetext[2]{Ellipticity: $1-b/a$, where $b$ is the semi-minor axis and $a$ is the semi-major axis.}

\end{minipage}
\end{table*}
%%%%%%%%%%%%%%%%%%%%%%%%%%%%%%%%%%%%%%%%%%%%%%%%%%%%%%%%%%%%%%%%%%%%%%%%%%%%%%%

%%%%%%%%%%%%%%%%%%%%%%%%%%%%%%%%%%%%%%%%%%%%%%%%%%%%%%%%%%%%%%%%%%%%%%%%%%%%%%%
% TABLE 2
\begin{table}
\centering
\caption{Log of WFC observations of the And\,I dwarf galaxy}
\begin{tabular}{cccccc}
	\hline\hline
 \noalign{\smallskip}
	Date                     &
	Epoch                    &
	Filter                   &
	$t_{\rm exp}$            &
        Airmass                  &
        seeing                  \\
        (y\,m\,d)                &
        &
        &
        (sec)                    &
        &
        (arcsec)               \\ 
 \noalign{\smallskip}       
	\hline
\noalign{\smallskip}
2016 02 09  &  2  &  $i$  &  540   &  1.475  &  1.35  \\
2016 06 14  &  3  &  $i$  &  555   &  1.568  &  1.68  \\
2016 08 10  &  4  &  $i$  &  555   &  1.077  &  1.39  \\
2016 08 12  &  4  &  $V$  &  735   &  1.015  &  1.30  \\
2016 10 20  &  5  &  $i$  &  555   &  1.205  &  1.48  \\
2016 10 20  &  5  &  $V$  &  735   &  1.286  &  1.53  \\
2017 01 29  &  6  &  $i$  &  555   &  2.425  &  1.77  \\
2017 08 01  &  7  &  $i$  &  555   &  1.018  &  1.18  \\ 
2017 08 01  &  7  &  $V$  &  735   &  1.014  &  1.25  \\
2017 09 01  &  8  &  $i$  &  555   &  1.084  &  1.25  \\ 
2017 09 01  &  8  &  $V$  &  735   &  1.051  &  1.34  \\
2017 10 06  &  9  &  $i$  &  555   &  1.278  &  1.44  \\
2017 10 08  &  9  &  $V$  &  735   &  1.014  &  1.21  \\ 
 \noalign{\smallskip}
\hline
\end{tabular}
\end{table}
%%%%%%%%%%%%%%%%%%%%%%%%%%%%%%%%%%%%%%%%%%%%%%%%%%%%%%%%%%%%%%%%%%%%%%%%%%%%%%%
\section{Survey design and Observations}

Over a period of three years (June 2015 to February 2018), we used the Wide Field Camera (WFC) to survey the majority of dwarf galaxies in the LG (Saremi et al.\ 2017). The WFC is an optical mosaic camera at the 2.5-m Isaac Newton Telescope (INT) of the Observatorio del Roque de los Muchachos (La Palma). It consists of four $2048\times4096$ CCDs, with a pixel size of $0\rlap{.}^{\prime\prime}33$ pixel$^{-1}$. The edge-to-edge limit of the mosaic, neglecting the $\sim1^\prime$ inter-chip spacing, is $34\rlap{.}^{\prime}2$.

Our sample includes 43 dSph, 6 dIrr, 6 dTrans and 4 globular clusters (GCs) all visible in the Northern hemisphere. Due to the time limitation for observation, we removed some of the dwarf galaxies which already had been studied with our method, such as NGC\,147 and NGC\,185 (Hamedani Golshan et al.\ 2017), and IC\,1613 (Hashemi et al.\ 2018). The distribution of our sample in the sky is shown in Fig.\ 1.

The main priority of this survey was the observation of the majority of Andromeda satellites because these are all accessible to a Northern hemisphere survey and provide an excellent sample due to homogeneity in distances, completeness, accuracy, and foreground contamination and extinction. They allow the study of a complete set of satellites of an L$_\star$ galaxy. Based on their estimated number of AGB stars, we further prioritised the remaining targets; some populous galaxies include IC\,10 (probably Andromeda's satellite; van den Bergh 2000; with $>10^4$ AGB stars) and Sextans\,A and B (isolated dwarfs with $>10^3$ AGB stars), although based on their heliocentric radial velocities, probably the last two are not LG members (van den Bergh 1999). In order to find out whether the Andromeda system could be considered as a universal template for galaxy evolution or it is just a particular case, the Milky Way satellites were also observed for comparison. Also, we included four distant GCs: Sextans\,C, Segue\,3, NGC\,2419 and Palomar\,4, to investigate the possibility of them being stripped nucleated dwarf galaxies. Although GCs are comparable in terms of luminosity and velocity dispersion with dwarf galaxies, they do not contain any dark matter (Pe\~narrubia et al.\ 2008; Tolstoy et al.\ 2009). Some of the observational properties of our sample targets are summarised in Table 1.

We aimed to monitor over ten epochs to identify LPVs and determine their amplitude and mean brightness. These epochs are spaced a few months apart, because they vary on timescales from $\sim60$ days (McDonald \& Zijlstra 2016) to $\sim700$ days (Vassiliadis \& Wood 1993) for AGB stars, and up to $\sim2000$ days for the longest-period supergiants (Samus et al.\ 2006). The first epoch began in June 2015, and the last one was completed in February 2018. Unfortunately, we lost most nights in February 2018 due to bad weather. The details of the observations for And\,I are listed in Table 2; details for the observations of the other targets can be found in an appendix.

According to the dimensions of our target galaxies and the field of view of WFC (about $34^\prime\times34^\prime$), each galaxy was observed in just one pointing and in the majority of cases centered on one of the four CCDs. To fill in the chip gaps and increase the signal-to-noise ratio, the objects were observed nine times with offsets of $30^{\prime\prime}$ between the pointings. Also, for the purpose of photometric calibration, standard star fields were observed each night with the same conditions as the targets.

Observations were taken in the WFC Sloan $i$ and Harris $V$ filters (except the first epoch in which the WFC RGO $I$ filter was used). We selected the $i$-band because the SEDs of cool evolved stars peak around 1 $\mu$m and the contrast between LPVs and other stars is enhanced. Also, the bolometric correction needed to determine the luminosity in this band is the smallest and the effects of attenuation by dust are minimal. To monitor variations in temperature -- and thus radius -- and more accurately model the SEDs, we also observed several times in the $V$-band.

We chose exposure times that yielded sufficient signal-to-noise ($S/N$) to detect small changes in magnitude at different epochs. The $i$-band amplitudes of pulsating AGB stars are $>0.1$ mag. Therefore we aimed for $S/N=10$ for the faintest stars, equivalent to the tip of the RGB. Going deeper would quickly lead to crowding-limited conditions. With such exposure times, depending upon the distance of the dwarf galaxies, this photometry is sufficiently deep to detect individual LPV stars (cf.\ Sec.\ 3.2).

%@@@@@@@@@@@@@@@@@@@@@@@@@@@@@@@@@@@@@@@@@@@@@@@@@@@@@@@@@@@@@@@@@@@@ section 3
%
\section{Data processing}

The images were processed using the {\sc theli} (Transforming HEavenly Light into Image) code, an image processing pipeline for optical images taken by multi-chip (mosaic) CCD cameras (Erben et al.\ 2005). It consists of several shell scripts that each perform a specific task and can run in parallel. The main steps include:
\begin{itemize}
\item{Separating the image files into frames of the individual chips (four CCDs in a WFC mosaic);}
\item{Removing instrumental signatures from the data: the electronic offset (bias), pixel response and instrumental throughput variations (flatfield) and interference in the back-illuminated thinned detector chip, especially prevalent at the reddest wavelengths which more closely match the physical thickness of the chip (fringe pattern);}
\item{Creating weight maps for individual frames based on the normalised flats. They can mask defects such as cosmic and hot pixels in the images;}
\item{Astrometric calibration to create a full astrometric solution taking into account the gaps between the chips and overlapping objects;}
\item{Subtracting the sky background from all frames;}
\item{Combining the images and creating a co-added image using a weighted mean 
method.}
\end{itemize}

The {\sc theli} pipeline is obtimised to perform precise astrometry and it is ideally suited to our goal. The {\sc daophot/allstar} software (Stetson 1987) was used to obtain photometry for all stars in our crowded stellar fields by employing a Point Spread Function (PSF) fitting method. Because some of the parameters (such as gain and readout noise) were different, we had to do photometry for each of the four CCDs separately. In most cases, the central chip covers the whole dwarf galaxy, so we decided to analyse this chip initially and leave the remaining chips for later.

After the initial object detection (from peaks above the noise), we selected 50--70 isolated PSF candidates in each frame to build a PSF model. Then {\sc allstar} subtracted all the stars from the image using PSF-fitting photometry along with the best current guesses of their positions and magnitudes. The {\sc allstar} photometry files of the individual images were aligned using {\sc daomaster}. The {\sc montage2} routine combined the individual images to create a master mosaic of each galaxy. A master catalogue of stars was used as input for the {\sc allframe} to perform simultaneous PSF-fitting photometry within each of the individual images (Stetson 1994).

%@@@@@@@@@@@@@@@@@@@@@@@@@@@@@@@@@@@@@@@@@@@@@@@@@@@@@@@@@@@@@@@@@@ section 3.1

\subsection{Calibration}

%%%%%%%%%%%%%%%%%%%%%%%%%%%%%%%%%%%%%%%%%%%%%%%%%%%%%%%%%%%%%%%%%%%%%%%%%%%%%%%
% FIGURE 2
\begin{figure}
\includegraphics[width=1.0\columnwidth]{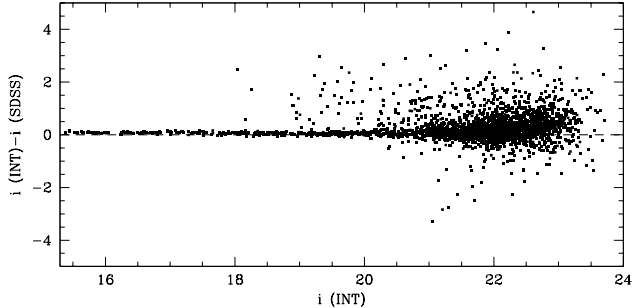}
\caption{Magnitude differences between INT catalogue and SDSS of And\,I, plotted against $i$ magnitude of our catalogue.}
\end{figure}
%%%%%%%%%%%%%%%%%%%%%%%%%%%%%%%%%%%%%%%%%%%%%%%%%%%%%%%%%%%%%%%%%%%%%%%%%%%%%%%

The calibration process was performed in several steps. First, aperture corrections to the PSF-fitting photometry were made as follows:
\begin{itemize}
\item{Selecting 30 stars with good photometry using the {\sc allframe} fitting parameters $\sigma$, $\chi$ (goodness-of-fit) and ``sharpness'' value as local 
standards;}
\item{Constructing growth-curves for each frame with the {\sc daogrow} routine (Stetson 1990) from which all stars were subtracted except the 30 selected stars;}
\item{Calculating the aperture corrections, i.e.\ the difference between the PSF-fitting and large-aperture magnitude of these stars with the {\sc collect} routine (Stetson 1993);}
\item{Adding the aperture corrections to each of the PSF-fitting magnitudes using the {\sc newtrial} routine (Stetson 1996);}
\end{itemize}

Next, in order to perform photometric calibration, we determined zero points for each frame based on the standard star field observations (for frames obtained on nights without standard star measurements, the average of zero points from other nights was used). We applied transformation equations derived from comparing Stetson's compilation of the Landolt standard stars with the corresponding Sloan Digital Sky Survey (SDSS) DR4 photometry (Jordi et al.\ 2006) to obtain accurate zero points in the Sloan $i$ filter. Airmass-dependent atmospheric extinction corrections were applied to adopt the extinction coefficients determined for La Palma (Garc\'{\i}a-Gil et al.\ 2010).

Finally, relative photometry between epochs was performed to correctly separate the variable from non-variable sources. For this, we selected approximately 1000 stars in common between all frames within the magnitude interval $i\in [18...21]$ mag. Then for each star, separately, we determined the deviation in each of the epochs with respect to the mean magnitude for that star on all epochs. The mean magnitudes were calculated weighting the individual measurements:
\begin{equation}
\langle m\rangle=\frac{\Sigma_{i=1}^n \frac{m_i}{\sigma_i^2}}{\Sigma_{i=1}^n \frac {1}{\sigma_i^2}} 
\end{equation}
where $\sigma_i$ is the estimated uncertainty of each magnitude determination (photometric errors). Then we averaged these deviations for each epoch. The result of the corrections, which are between $-0.0201$ to 0.0252 mag (but usually much smaller), was applied to frames.

To estimate the accuracy of calibration, we cross-correlated the results with the SDSS catalogue. The matches were obtained by performing search iterations using growing search radii in steps of $0\rlap{.}^{\prime\prime}1$ out to $1^{\prime\prime}$, on a first-encountered first-associated basis but after ordering the principal photometry in order of diminishing brightness (to avoid rare bright stars being erroneously associated with any of the much larger numbers of faint stars). For example, the result of cross-correlation of the And\,I catalogue, as shown in Fig.\ 2, is consistent within good agreement with the SDSS catalogue within our desired range (see below).

%@@@@@@@@@@@@@@@@@@@@@@@@@@@@@@@@@@@@@@@@@@@@@@@@@@@@@@@@@@@@@@@@@@ section 3.2

\subsection{Quality assessment}

%%%%%%%%%%%%%%%%%%%%%%%%%%%%%%%%%%%%%%%%%%%%%%%%%%%%%%%%%%%%%%%%%%%%%%%%%%%%%%%
% FIGURE 3
\begin{figure}
\includegraphics[width=1.0\columnwidth]{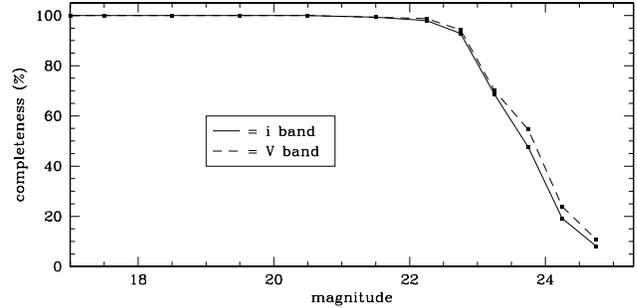}
\caption{Completeness as a function of $i$-band (solid line) and $V$-band (dashed line) magnitude for And\,I data. The photometry is better than 98\% complete at $i\sim22$ mag, dropping to below 50\% at $i=23.6$ mag.}
\end{figure}
%%%%%%%%%%%%%%%%%%%%%%%%%%%%%%%%%%%%%%%%%%%%%%%%%%%%%%%%%%%%%%%%%%%%%%%%%%%%%%%
%%%%%%%%%%%%%%%%%%%%%%%%%%%%%%%%%%%%%%%%%%%%%%%%%%%%%%%%%%%%%%%%%%%%%%%%%%%%%%%
% FIGURE 4
\begin{figure}
\includegraphics[width=1.0\columnwidth]{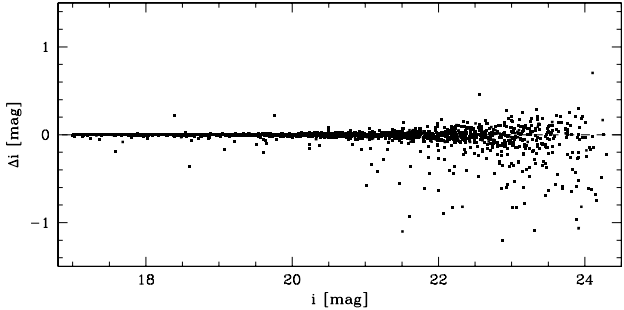}
\caption{The difference between the input stellar magnitudes and the recovered stellar magnitudes from the artificial star tests. For stars brighter than $i\sim22$ mag the scatter is $<0.1$ mag.}
\end{figure}
%%%%%%%%%%%%%%%%%%%%%%%%%%%%%%%%%%%%%%%%%%%%%%%%%%%%%%%%%%%%%%%%%%%%%%%%%%%%%%%
%%%%%%%%%%%%%%%%%%%%%%%%%%%%%%%%%%%%%%%%%%%%%%%%%%%%%%%%%%%%%%%%%%%%%%%%%%%%%%%
% FIGURE 5
%
\begin{figure}
\includegraphics[width=1.0\columnwidth]{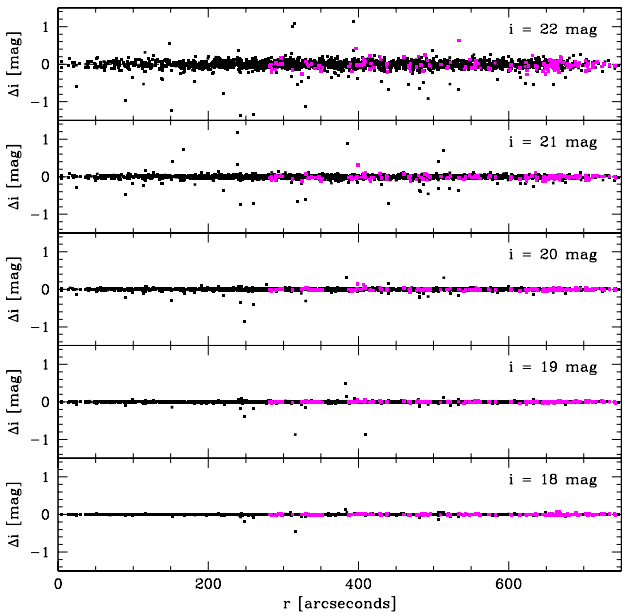}
\caption{Difference between input magnitude and recovered magnitude (from an i-band image) {\it vs}.\ distance from the centre of And\,I, for five values of the input magnitude. Stars observed near the edges of the frames are highlighted in magenta.}
\end{figure}
%%%%%%%%%%%%%%%%%%%%%%%%%%%%%%%%%%%%%%%%%%%%%%%%%%%%%%%%%%%%%%%%%%%%%%%%%%%%%%%

To assess the survey completeness, the {\sc addstar} task of the {\sc daophot} package (Stetson 1987) was performed. By adding synthetic stars to an image and applying all steps of photometry, we estimated the star-finding efficiency and the photometric accuracy. This was done by comparing the output data for these stars to what had been put in. We added 300 artificial stars in each of seven trials (to avoid crowdedness) to the master mosaic of And\,I and two individual frames of it in $i$ and $V$-bands (as an example) in 1-mag bins starting from $i=17$ mag until $i=25$ mag (for $V$ the same as $i$). Having done the {\sc daophot/allstar/allframe} procedure on these new frames, we used {\sc daomaster} to find the fraction of recovered stars. As shown in Fig.\ 3, the photometry is better than 98\% complete at $i\sim22$ mag, which is fainter than the RGB tip (cf.\ Sec.\ 5.1). The completeness limit drops below 50\% at $i=23.6$ mag. The $V$-band reaches similar completeness levels but at $\sim0.1$ magnitude fainter. To check the crowding effect in the centre of the galaxy, we also calculated the completeness limit inside the half-light radius region only; the results were quite similar to the previous one.

The mean difference between the magnitudes of the added and recovered stars (Fig.\ 4) shows a small scatter which increases with magnitude. For stars brighter than the 98\% completeness limit (to $i\sim22$ mag) it is $|\Delta i|<0.1$ mag. Since our target candidates typically have $i<22$ mag (except for rare, heavily dust-enshrouded stars that also tend to have the largest amplitudes), the photometry is deep and accurate enough to find essentially all of the LPV candidates.

We also examined the difference between the input magnitude and the recovered magnitude {\it vs}.\ distance from the centre of the galaxy, for varying magnitudes in a 0.07 square-degree region that includes the galaxy centre. To that aim, we added 300 stars in each of seven trials with magnitudes $i\in (18,19,20,21,22)$ mag to one of And\,I's individual frames and then we repeated the photometry for this frame. As one can see in Fig.\ 5, the difference in magnitude is minimal and only slightly increases for $i=22$ mag. Stars that are located near the edges of the frame (highlighted in magenta) show the same behaviour as the stars elsewhere in the image. So we conclude that the photometry is accurate everywhere in the frame uniformly.

%@@@@@@@@@@@@@@@@@@@@@@@@@@@@@@@@@@@@@@@@@@@@@@@@@@@@@@@@@@@@@@@@@@@@ section 4
%
\section{Identification of LPVs}

%%%%%%%%%%%%%%%%%%%%%%%%%%%%%%%%%%%%%%%%%%%%%%%%%%%%%%%%%%%%%%%%%%%%%%%%%%%%%%%
% FIGURE 6
\begin{figure}
\includegraphics[width=1.0\columnwidth]{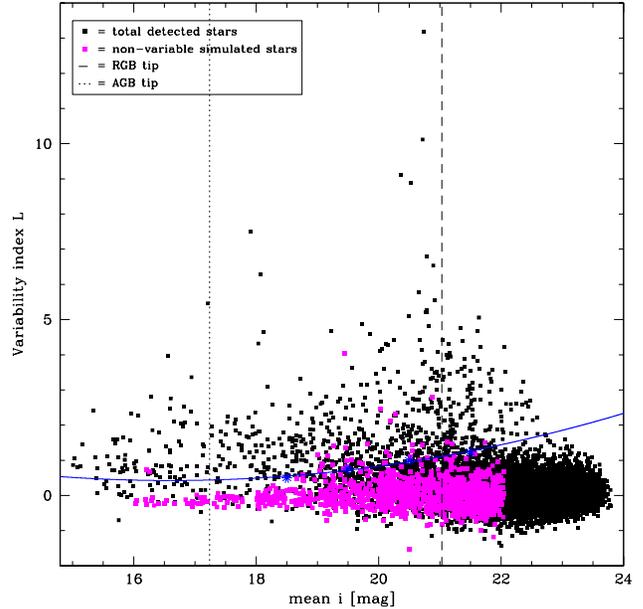}
\caption{Variability index $L$ {\it vs}.\ $i$-band magnitude for all stars in the catalogue of And\,I. The vertical lines represent the tips of the RGB and AGB (cf.\ Sec.\ 5.1). Simulated non-variable stars (magenta points) are overplotted to decide the detection limit for the variable stars. The blue curve indicates our threshold for identifying variable stars (see Fig.\ 7). With this selection criterion, only $2.5$\% of simulated stars with constant magnitude are found above the line.}
\end{figure}
%%%%%%%%%%%%%%%%%%%%%%%%%%%%%%%%%%%%%%%%%%%%%%%%%%%%%%%%%%%%%%%%%%%%%%%%%%%%%%%

The method used in this project to find LPVs is similar to the procedure described by Welch \& Stetson (1993) and Stetson (1996). In this way, first, the weighted mean magnitudes, $\langle m\rangle$, for all stars in the two filters $i$ and $V$ are calculated according to equation 1. The normalized magnitude residuals, which are scaled by the photometric errors, are defined by
\begin{equation}
\delta_i={\sqrt{\frac {n}{n-1}}}{\frac{m_i-\langle m\rangle}{\sigma_i}}
\end{equation}
where $n$ is the total number of observations contributing to the mean. To calculate variability indices, we paired those observations in the $i$ and $V$ bands obtained within a timespan less than the shortest period expected for the LPVs (60 days) and assigned a weight $w_K$ to each pair of them. If within a pair of observations only one measurement for a star was available, $w_k$ was set to 0.5. Then the $J$ index is calculated:
\begin{equation}
J=\frac{\Sigma_{k=1}^N w_k\ {\rm sign}(P_k)\sqrt{|P_k|}}{\Sigma_{k=1}^N w_k}.
\end{equation}
where $N$ is the total number of observations and
\begin{equation}
P_k=\left\{
\begin{array}{lll}
(\delta_i\delta_j)_k & {\rm if} & i\neq j \\
\delta_i^2-1 & {\rm if} & i=j \\
\end{array}
\right.
\end{equation}  
The $J$ index has a large positive value for variable stars and tends to zero for data containing random noise only.

Based on the shape of light-curve, the Kurtosis index $K$ is determined:
\begin{equation}
K=\frac{\frac{1}{N} \Sigma_{i=1}^N |\delta_i|}{\sqrt{\frac{1}{N}\Sigma_{i=1}^N\delta_i^2}}.
\end{equation}
with $K=0.9$ for a sinusoidal light variation, $K=0.798$ for a Gaussian magnitude distribution where random measurement errors dominate over physical variation, and $K\rightarrow0$ for data affected by a single outlier.

Finally, we calculated the variability index $L$:
\begin{equation}
L=\frac{J\times K}{0.798} \frac {\Sigma_{i=1}^N w_i}{w_{\rm all}},
\end{equation}
based on variability indices $J$ and $K$ and an extra weight assigned to the stars with the highest number of measurements. Fig.\ 6 shows how the variability index $L$ varies with $i$-band magnitude for all stars in the catalogue of And\,I. As one can see the plot reveals an excess of stars with larger than usual $L$ between $i\sim19$--22 mag; these are likely AGB stars with Mira-type variability. It is expected that most of the bright stars between $i\sim15$--18 mag will be foreground stars because the presence of RSGs in And\,I is unlikely (cf.\ Sec.\ 4.1).
 
To determine the threshold that separates variable from non-variable sources, several tests were performed. First, histograms of the variability index $L$ for several $i$-band magnitude intervals in the range 18--22 mag were plotted (Fig.\ 7). To estimate an $L$ threshold, we mirrored the negative part of these histograms, as LPVs are not expected among sources with $L<0$. Then, a Gaussian function was fitted to the negative part and its mirror. The Gaussian distribution is a perfect fit at low values of $L$ while it departs from the histograms for larger (positive) values. To determine the point at which the likelihood not to be part of the Gaussian distribution exceeds 90\% (i.e.\ only 10\% contamination by non-variables), we estimated the corresponding $L$ threshold value for each brightness interval (labeled on Fig.\ 7). As expected from Fig.\ 6, a magnitude-dependent threshold resulted. By fitting a polynomial function to these threshold points in Fig.\ 6, we identified 470 LPV candidates in the region of CCD 4 of WFC -- which are clearly not all members of And\,I.

%%%%%%%%%%%%%%%%%%%%%%%%%%%%%%%%%%%%%%%%%%%%%%%%%%%%%%%%%%%%%%%%%%%%%%%%%%%%%%%
% FIGURE 7
\begin{figure}
\includegraphics[width=1.0\columnwidth]{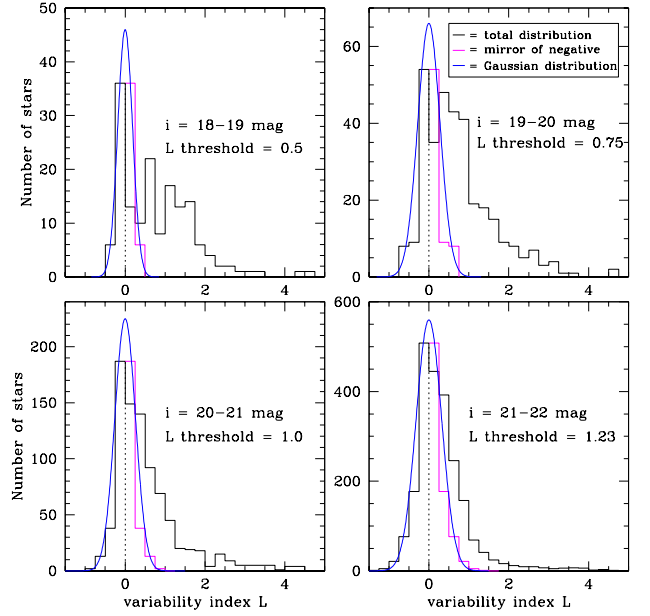}
\caption{Histograms of the variability index $L$, for several $i$-band magnitude bins. The negative part of each histogram is mirrored to facilitate a Gaussian fit representing the expected distribution of non-variables. The optimal variability index thresholds are labeled on the panels.}
\end{figure}
%%%%%%%%%%%%%%%%%%%%%%%%%%%%%%%%%%%%%%%%%%%%%%%%%%%%%%%%%%%%%%%%%%%%%%%%%%%%%%%
 
To further assess the validity of the choice of variability index thresholds, a simulation was made of stars with constant magnitude light-curves with the {\sc addstar} task of the {\sc daophot} package (Stetson 1987). We added 1500 synthetic stars within the magnitude interval $i\in [16...22]$ mag in six trials on all frames and we did photometry and calculated variability indices in the same way as before. A similar plot of variability index {\it vs}.\ magnitude is used to decide on the separation limit between the variable and non-variable stars (overplotted on Fig.\ 6). By setting the curve fit estimated previously, only $2.5$\% of stars with constant magnitude from these simulations are found above this limit, which instills confidence in the reliability of our selection.

As a preliminary check on the validity of candidate variables, we inspected the images of all of these sources by eye and divided them into two categories; among 470 stars which were selected as candidate variables in the And\,I master catalogue, 251 stars (class 1) had a peak distinct from their neighbours, displayed with green squares on the colour--magnitude diagram (CMD) in Fig.\ 8. The others (class 2) were involved in the profile of a brighter star and/or hopelessly confused with multiple neighbours and did not look like a true variable star (shown with red squares in Fig.\ 8). Highlighted by yellow triangles are unreliable candidates with only five or fewer measurements. The estimated completeness limit (cf.\ Sec.\ 3.2) is shown with the red dotted line in Fig.\ 8; the horizontal dotted lines represent the tips of the RGB and AGB (cf.\ Sec.\ 5.1).

%%%%%%%%%%%%%%%%%%%%%%%%%%%%%%%%%%%%%%%%%%%%%%%%%%%%%%%%%%%%%%%%%%%%%%%%%%%%%%%
% FIGURE 8
\begin{figure}
\includegraphics[width=1.0\columnwidth]{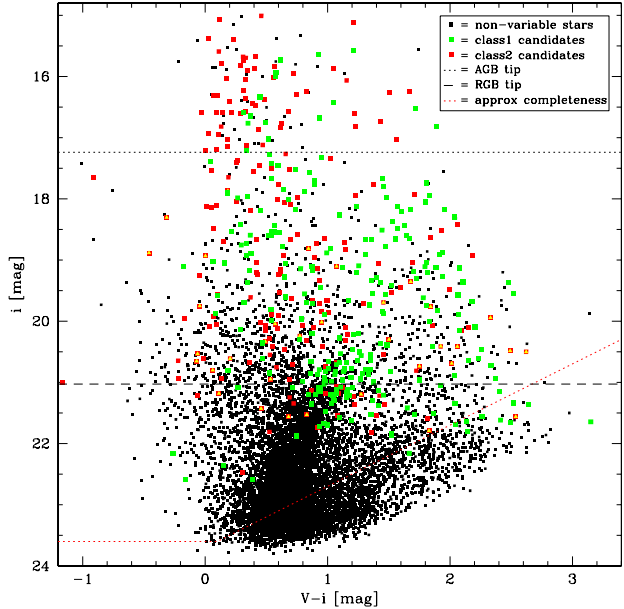}
\caption{CMD of And\,I in the $i$-band {\it vs}.\ $V-i$ colour with indicated the candidate variables according to two categories. The green squares are more reliable candidates that have an optical profile with a peak distinct from their neighbours (class 1). Unreliable candidate variables are shown with red squares (class 2). Stars with only five or fewer measurements are highlighted with yellow triangles.}
\end{figure}
%%%%%%%%%%%%%%%%%%%%%%%%%%%%%%%%%%%%%%%%%%%%%%%%%%%%%%%%%%%%%%%%%%%%%%%%%%%%%%%

We plotted light-curves of three LPV candidates with relatively large amplitudes along with a non-variable star for comparison (Fig.\ 9). The bottom light-curve is related to a star which is also found to be variable at mid-IR wavelengths in the {\it Spitzer} Space Telescope monitoring survey DUSTiNGS (Boyer et al.\ 2015b; cf.\ Sec.\ 5.3.1). It is a very dusty AGB star and therefore not seen in the $V$ band, so its colour is evaluated on the basis of our completeness limit in the $V$ band (23.7 mag; $V-i>2.04$ mag).

%%%%%%%%%%%%%%%%%%%%%%%%%%%%%%%%%%%%%%%%%%%%%%%%%%%%%%%%%%%%%%%%%%%%%%%%%%%%%%%
% FIGURE 9
\begin{figure}
\includegraphics[width=1.0\columnwidth]{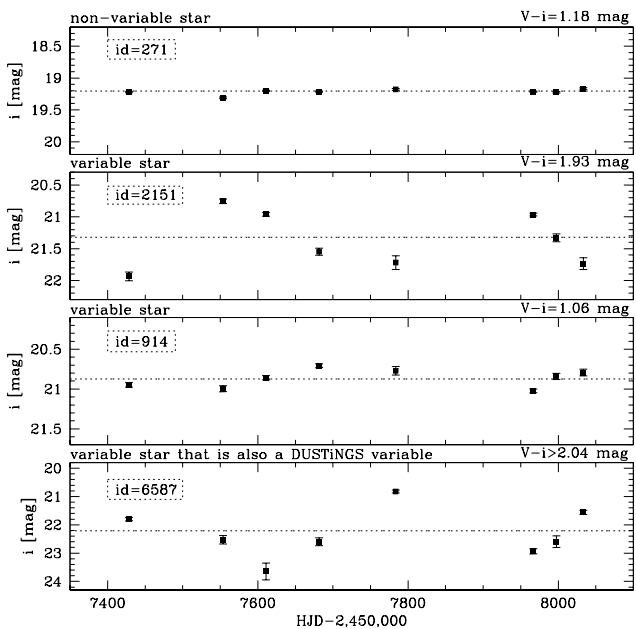}
\caption{Example light-curves of three LPV candidates with relatively large amplitudes with a non-variable star in the top panel for comparison. The variable star \#6587 in the bottom panel is in common with the DUSTiNGS catalogue of variable {\it Spitzer} sources (Boyer et al.\ 2015b; cf.\ Sec.\ 5.3.1).}
\end{figure}
%%%%%%%%%%%%%%%%%%%%%%%%%%%%%%%%%%%%%%%%%%%%%%%%%%%%%%%%%%%%%%%%%%%%%%%%%%%%%%%

Also, we investigated the colour changes of LPV candidates one by one. It is expected that LPVs become redder when they become optically fainter. Most of the LPV candidates (class 1) within two half-light radii from the centre of And\,I had the expected colour changes and only a few ones displayed an opposite behaviour; however, the latter had small amplitudes, so this can be attributed to scatter in colour from photometric uncertainties. Outside two half-light radii from the centre of And\,I, five LPV candidates had an unexpected behaviour and were removed from our LPV candidates catalogue. Fig.\ 10 shows two examples of LPV candidates with expected and unexpected colour changes.

%%%%%%%%%%%%%%%%%%%%%%%%%%%%%%%%%%%%%%%%%%%%%%%%%%%%%%%%%%%%%%%%%%%%%%%%%%%%%%%
% FIGURE 10
\begin{figure}
\includegraphics[width=1.0\columnwidth]{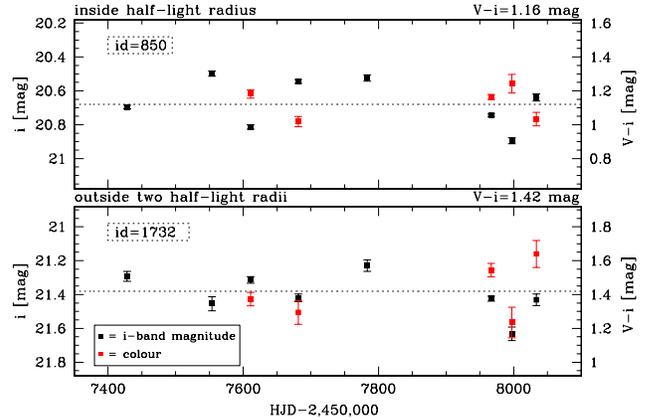}
\caption{Example light-curves of two LPV candidates along with their colour changes. The variable star \#850 in the top panel exhibits the expected colour changes, becoming redder when becoming optically fainter. The bottom panel shows a LPV candidate that does not display thee expected behaviour in colour.}
\end{figure}
%%%%%%%%%%%%%%%%%%%%%%%%%%%%%%%%%%%%%%%%%%%%%%%%%%%%%%%%%%%%%%%%%%%%%%%%%%%%%%%

It should be noted that a number of images taken on June 14th 2016 and January 29th 2017 do not have excellent quality in sharpness values. This is probably related to the bad seeing and photometric conditions on those nights. Fig.\ 11 shows the sharpness values, {\it vs}.\ $i$-band magnitude for these frames as well as the September 1st 2017 frame for comparison; the candidate variables are highlighted on these plots. The increased scatter in these two frames is obvious to see, and especially the frame of June 2016 is poor, so we decided to decrease the weight $w_K$ (see equation 3) for these frames compared to other frames.

%%%%%%%%%%%%%%%%%%%%%%%%%%%%%%%%%%%%%%%%%%%%%%%%%%%%%%%%%%%%%%%%%%%%%%%%%%%%%%%
% FIGURE 11
\begin{figure}
\includegraphics[width=1.0\columnwidth]{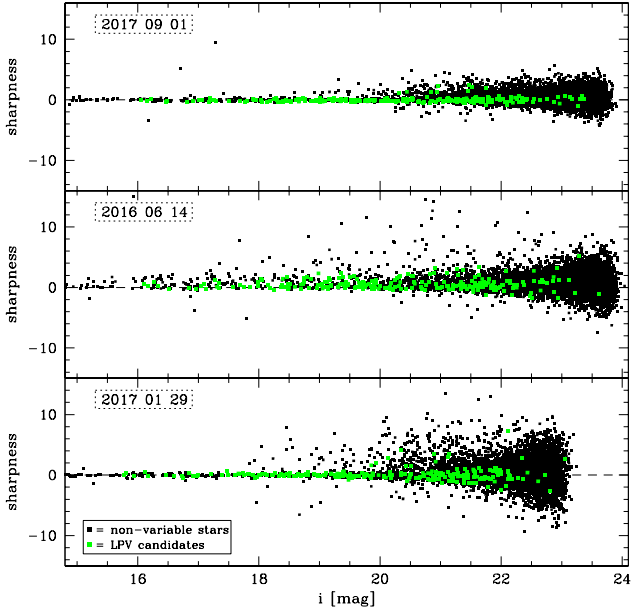}
\caption{The sharpness values, {\it vs}.\ $i$-band magnitude for three of the individual frames. The candidate variables are highlighted in green.}
\end{figure}
%%%%%%%%%%%%%%%%%%%%%%%%%%%%%%%%%%%%%%%%%%%%%%%%%%%%%%%%%%%%%%%%%%%%%%%%%%%%%%%

%@@@@@@@@@@@@@@@@@@@@@@@@@@@@@@@@@@@@@@@@@@@@@@@@@@@@@@@@@@@@@@@ subsection 4.1

\subsection{Contaminations}

%%%%%%%%%%%%%%%%%%%%%%%%%%%%%%%%%%%%%%%%%%%%%%%%%%%%%%%%%%%%%%%%%%%%%%%%%%%%%%%
% FIGURE 12
%
\begin{figure}
\includegraphics[width=1.0\columnwidth]{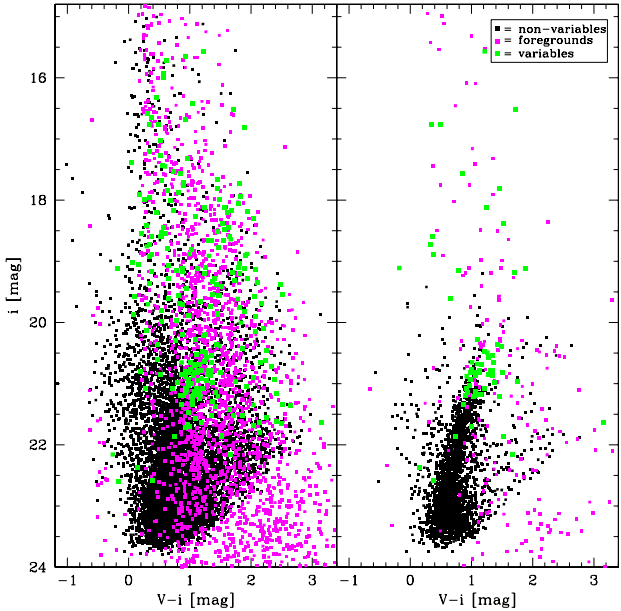}
\caption{A simulation with {\sc trilegal} (Girardi et al.\ 2005) of the expected contamination by foreground stars (in magenta). Our candidate variables are highlighted in green. The left panel concerns a 0.07 deg$^2$ field while the right panel concerns a 0.007 deg$^2$ field centred on And\,I corresponding to the half-light radius of the galaxy.}
\end{figure}
%%%%%%%%%%%%%%%%%%%%%%%%%%%%%%%%%%%%%%%%%%%%%%%%%%%%%%%%%%%%%%%%%%%%%%%%%%%%%%%

We present an estimate of the foreground stars in the direction of And\,I in Fig.\ 12, simulated with the {\sc trilegal} stellar population synthesis code (Girardi et al.\ 2005) via its web interface. Two different sizes for the field were considered; 0.07 deg$^2$ (about the size of the entire CCD 4 of WFC; left panel) and 0.007 deg$^2$ (the half-light radius of And\,I; right panel), in the direction ($l=121\rlap{.}^{\circ}68$, $b=-24\rlap{.}^{\circ}82$). Our candidate variables are highlighted in green on these plots. As one can see the area in the CMD that we are concerned with is contaminated by foreground stars; so the CMDs do not cleanly show the population of AGB stars and thus the SFH alone. However, the distributions over {\it LPVs} can be used to obtain the SFH relatively unaffected by the foreground contamination, because LPVs are relatively rare, especially in directions away from the Galactic plane (and the foreground contamination of the AGB at the distance of And\,I is comprised of non-AGB stars that are less variable).

%%%%%%%%%%%%%%%%%%%%%%%%%%%%%%%%%%%%%%%%%%%%%%%%%%%%%%%%%%%%%%%%%%%%%%%%%%%%%%%
% FIGURE 13
%
\begin{figure}
\includegraphics[width=1.0\columnwidth]{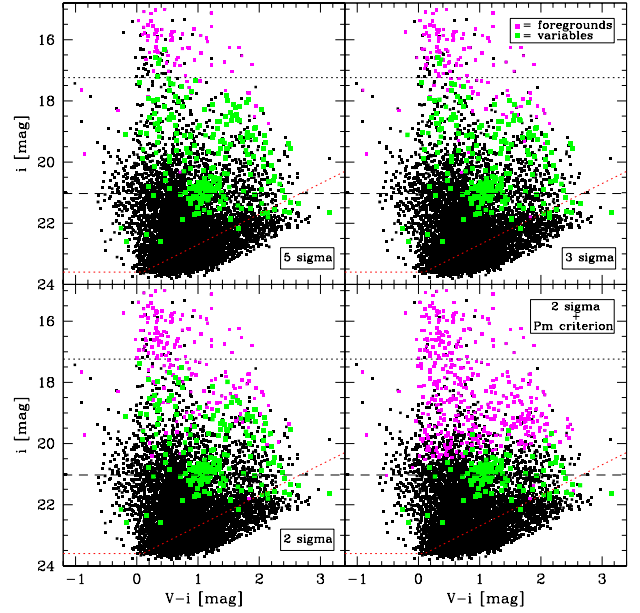}
\caption{Estimates of foreground stars based on {\it Gaia} DR2 data. By applying different criteria on the parallax and proper motion, the number of foreground stars (in magenta) and so the number of candidate variables (in green) change.}
\end{figure}
%%%%%%%%%%%%%%%%%%%%%%%%%%%%%%%%%%%%%%%%%%%%%%%%%%%%%%%%%%%%%%%%%%%%%%%%%%%%%%%

For a more precise and reliable estimate of foreground contamination, the master catalogue of And\,I was cross-correlated (as described in Sec.\ 3.1) with {\it Gaia} Data Release 2 (DR2; Gaia Collaboration 2018). Then the foreground stars were separated based on their parallaxes. First, we only selected as foreground those stars with parallax measurements that are significant at the $5 \sigma$ level -- these all greatly exceed what would be expected at the distance of And\,I. The result is shown in the top-left panel of Fig.\ 13 as a CMD of detected sources in which are highlighted the foreground stars in magenta and LPV candidates in green. In this way, from 10\,243 stars of our catalogue of the entire CCD 4 of WFC, only 81 stars are identified as foreground stars within the magnitude interval $i=15$--20 mag, among which 16 have a variability index above the threshold.

Based on the {\sc trilegal} simulation, the number of expected foreground stars within the same magnitude interval is 512, i.e.\ much larger. Therefore, we cross-correlated our catalogue with more uncertain {\it Gaia} DR2 data to get closer to the number predicted by the simulation. Cross-correlation with $3 \sigma$ and $2 \sigma$ criteria on the parallaxes resulted in 129 and 187 identified foreground stars shown in the top-right and bottom-left panels of Fig.\ 13, respectively. While the number of foreground stars increased, it is still far from the expected number in the {\sc trilegal} simulation. It seems therefore that selection on the basis of parallax measurement can only remove a (sizeable) minority of foreground sources.

There are two clear areas in the CMD where candidate variable stars appear, viz.\ between $i\sim17.5$--19.5 mag and colours of $V-i\sim0$--1 mag and $V-i\sim1$--2 mag. This should be further investigated because the presence of LPVs with such high luminosities constrains the young stellar population with an age of below 100 million years, even though such a population is not much expected for a dSph galaxy (e.g., Weisz et al.\ 2014). Also, comparison with the {\sc trilegal} simulation shows contamination in these areas bent towards redder colours, so it is possible that the candidates with $V-i\sim0$--1 mag are not in And\,I. They may be contaminated by stars in the Andromeda galaxy (M\,31).

Finally, we used proper motion (PM) measurements of {\it Gaia} DR2 along with a $2 \sigma$ level threshold criterion for the parallaxes (bottom-right panel of Fig.\ 13). We selected a criterion of $\sqrt{(\mu_{\alpha})^2+(\mu_{\delta})^2}>0.28$ mas yr$^{-1}$ $+2.0\ error$ for PM, similar to the one described by van der Marel et al.\ (2019). Stars were considered foreground objects if they satisfied {\it either} of these criteria. In this way, 421 stars of our catalogue were identified as foreground stars, i.e.\ much closer to our expectations, and hence all of the unexpected (luminous) LPV candidates were removed. With this selection, from 251 early LPV candidates of class 1, 128 were considered as foreground objects. Unfortunately, faint stars with $i\gtrsim 20$ mag are not very well characterised in the {\it Gaia} catalogue (it is essentially complete to $G=17$ mag; Gaia Collaboration 2018).

The {\sc trilegal} simulation only accounts for Milky Way foregrounds so it can not estimate the total of contaminations. It is especially evident in the left panel of the Fig.\ 12 in a region with $V-i \lesssim 0.6$ and $21\lesssim i \lesssim22.5$ mag. To estimate the background contamination, we used the part of the CCD that lies outside two half-light radii of And\,I. Fig.\ 14 presents CMDs of two regions with approximately the same area, one within the two half-light radii of And\,I and another outside of three half-light radii (respectively, left and right panels). By comparing the two CMDs, we can conclude that about 35\% identified sources and 28\% LPV candidates within two half-light radii of And\,I are contaminated. They may be stars in the M\,31 and/or background active galactic nuclei (AGN) and/or foregrounds that are below the limit of completeness of Gaia. The absence of a clear, curved RGB in the control field (right panel of Fig.\ 14) suggests that contamination from M\,31 is negligible and that most of the contamination comes from Galactic foreground. Contamination from background galaxies/AGN would affect mostly the region around $V-i\sim 2$ mag and $i\gtrsim 21.5$ mag, not much affecting the AGB (or RGB) portion of the And\,I CMD.

%%%%%%%%%%%%%%%%%%%%%%%%%%%%%%%%%%%%%%%%%%%%%%%%%%%%%%%%%%%%%%%%%%%%%%%%%%%%%%%
% FIGURE 14
%
\begin{figure}
\includegraphics[width=1.0\columnwidth]{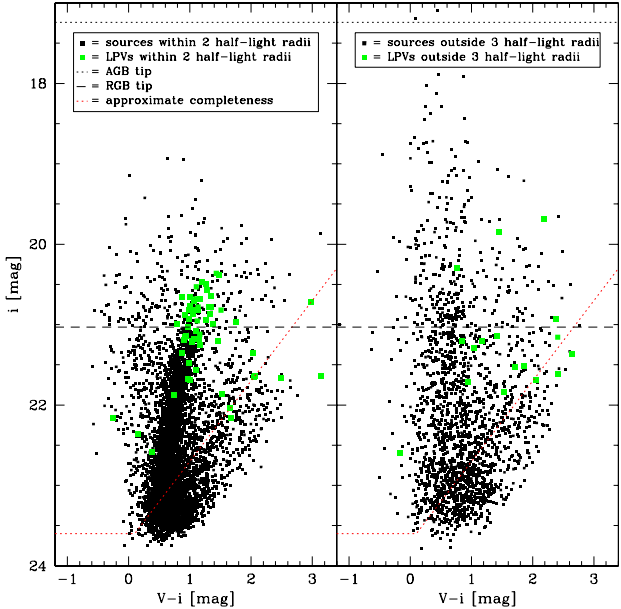}
\caption{The estimated background contamination with using CMD of And\,I in the $i$ band {\it vs}.\ $V-i$ colour showing our identified LPV candidates in green for two regions with the same area within two half-light radii (left panel) and outside three half-light radii (right panel).}
\end{figure}
%%%%%%%%%%%%%%%%%%%%%%%%%%%%%%%%%%%%%%%%%%%%%%%%%%%%%%%%%%%%%%%%%%%%%%%%%%%%%%%

%%%%%%%%%%%%%%%%%%%%%%%%%%%%%%%%%%%%%%%%%%%%%%%%%%%%%%%%%%%%%%%%%%%%%%%%%%%%%%%
% FIGURE 15
%
\begin{figure}
\includegraphics[width=1.0\columnwidth]{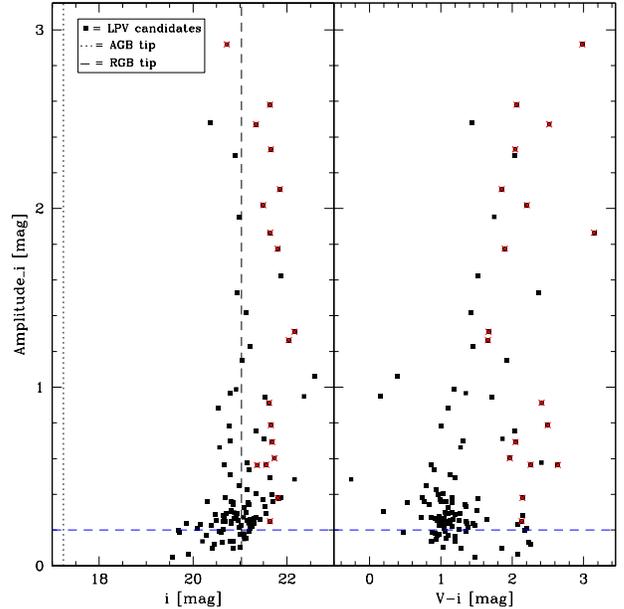}
\caption{The estimated amplitude of variability {\it vs}.\ $i$-band magnitude (left panel) and $V-i$ colour (right panel). Crosses indicate LPV candidates that were detected only twice or fewer times in the $V$-band, and so their colours are calculated based on the completeness limit. The dotted vertical lines represent the tips of the RGB and AGB. Candidate variables with $\Delta i <0.2$ mag are less reliable as LPVs.}
\end{figure}
%%%%%%%%%%%%%%%%%%%%%%%%%%%%%%%%%%%%%%%%%%%%%%%%%%%%%%%%%%%%%%%%%%%%%%%%%%%%%%%
%@@@@@@@@@@@@@@@@@@@@@@@@@@@@@@@@@@@@@@@@@@@@@@@@@@@@@@@@@@@@@@@ subsection 4.2
\subsection{Amplitudes of variability}

We estimated the amplitude of variability by assuming a sinusoidal light-curve shape. By comparing the standard deviation in the magnitudes to that expected for a completely random sampling of a sinusoidal variation (0.701), one can recover the amplitude from
\begin{equation}
Amplitude=2\times\sigma/0.701,
\end{equation}
which is defined as the difference between the minimum and maximum brightness. For identified candidate variables, we show the $i$-band amplitude of variability {\it vs}.\ $i$-band magnitude in the left panel of Fig.\ 15. The dotted lines represent the tips of the RGB and AGB (cf.\ Sec.\ 5.1). The amplitudes stay below $\Delta i\sim3$ mag and generally $\Delta i<1.5$ mag. This plot shows while the brightness increases, the amplitude of variability decreases, which is a known fact for populations of LPVs (Wood et al.\ 1992; Wood 1998; Whitelock et al.\ 2003; van Loon et al.\ 2008). Among our candidates, the large-amplitude variables fall in the category of Miras. Miras have $\Delta V>2.5$ mag and $\Delta i>1.5$ mag by the classical definition (though some stars with smaller amplitudes are similar to Miras; Feast \& Whitelock 2014).

Among very dusty AGB stars, though, stars with redder colours have the largest amplitudes (Wood et al.\ 1992; McDonald \& Zijlstra 2016; McDonald \& Trabucchi 2019); these are often also relatively luminous. This tendency is seen in the right panel of Fig.\ 15, in which the $i$-band amplitude of variability is plotted {\it vs}.\ $V-i$ colour. Some LPV candidates (highlighted by red crosses) are detected only once or twice in the $V$-band (or not at all); their colours are estimated on the basis of the completeness limit (cf.\ Sec.\ 3.2). Their large amplitudes and red colours make them prime candidates for being dusty AGB stars.

There is a population of candidate variables with $\Delta i <0.2$ mag, but such small amplitudes must be regarded with a great deal of caution. We thus limited our LPV candidates catalogue to those stars with variability amplitudes larger than 0.2 mag in the $i$-band.

%@@@@@@@@@@@@@@@@@@@@@@@@@@@@@@@@@@@@@@@@@@@@@@@@@@@@@@@@@@@@@@@ subsection 4.3

%%%%%%%%%%%%%%%%%%%%%%%%%%%%%%%%%%%%%%%%%%%%%%%%%%%%%%%%%%%%%%%%%%%%%%%%%%%%%%%
% FIGURE 16
\begin{figure}
\centering
\includegraphics[width=1.0\columnwidth]{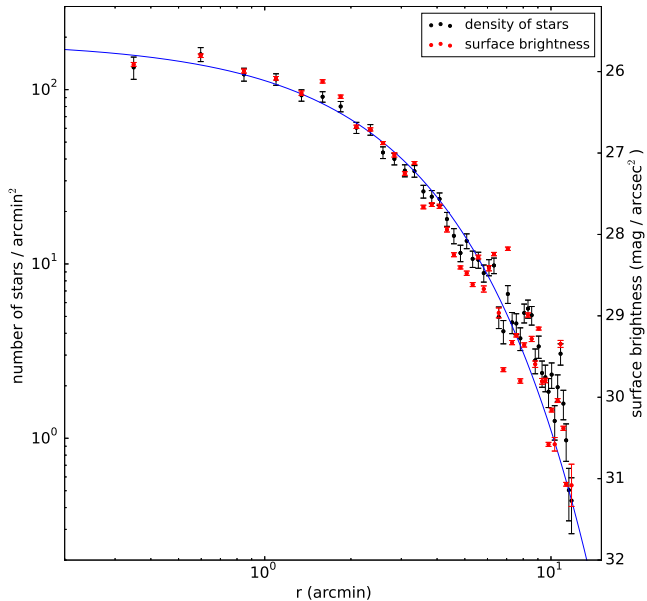}
\caption{The logarithmic form of the stellar number density (black points) as a function of radial distance to the centre of And\,I. Error bars take into account the Poisson uncertainty in the counts. On the right vertical axis, the surface brightness (red points) is constructed based on the integrated flux of the identified stars. The blue solid curve represents the best exponential fit to the data. We estimated a half-light radius of $3.2\pm0.3$ arcmin.}
\end{figure}
%%%%%%%%%%%%%%%%%%%%%%%%%%%%%%%%%%%%%%%%%%%%%%%%%%%%%%%%%%%%%%%%%%%%%%%%%%%%%%%

\subsection{Spatial location of candidate variables}

Caldwell et al.\ (1992) derived some of the structural parameters of And\,I using the integrated light of this galaxy; they reported 2.5 arcmin for the half-light radius of And\,I. Some of the next estimates were $3.1\pm0.3$ and $3.9\pm0.1$ arcmin by McConnachie (2012) and Martin et al.\ (2016), respectively. 
Here, we considered both the stellar number density and corresponding surface brightness determined in elliptical annuli as a function of radial distance to the centre of And\,I (Fig.\ 16). The surface brightness was derived from adding the fluxes of the identified stars. We fitted the radial profile with an exponential law and thus obtained a half-light radius of $3.2\pm0.3$ arcmin.

Fig.\ 17 presents the master image of CCD 4 of WFC, with And\,I located near its centre. Green circles represent the location of our identified LPV candidates and the half-light radius of And\,I is marked with a magenta ellipse. 
The density of variable stars within the ellipse, 1.27 arcmin$^{-2}$, is much larger than outside of it, 0.29 arcmin$^{-2}$, confirming that we have found LPV candidates belonging to And\,I. LPV candidates at distances greater than two half-light radii, shown as cyan circles, may be contaminated by stars in M\,31, Milky Way disc/halo, and by background galaxies; the majority of these are not expected to be members of And\,I. Based on the density of these candidates (0.24 arcmin$^{-2}$), we may expect 22 among the 59 identified LPV candidates within two half-light radii from the centre of And\,I to be spurious (cf.\ Sec.\ 4.1).

%%%%%%%%%%%%%%%%%%%%%%%%%%%%%%%%%%%%%%%%%%%%%%%%%%%%%%%%%%%%%%%%%%%%%%%%%%%%%%%
% FIGURE 17
\begin{figure*}
\centering
\includegraphics[width=19cm]{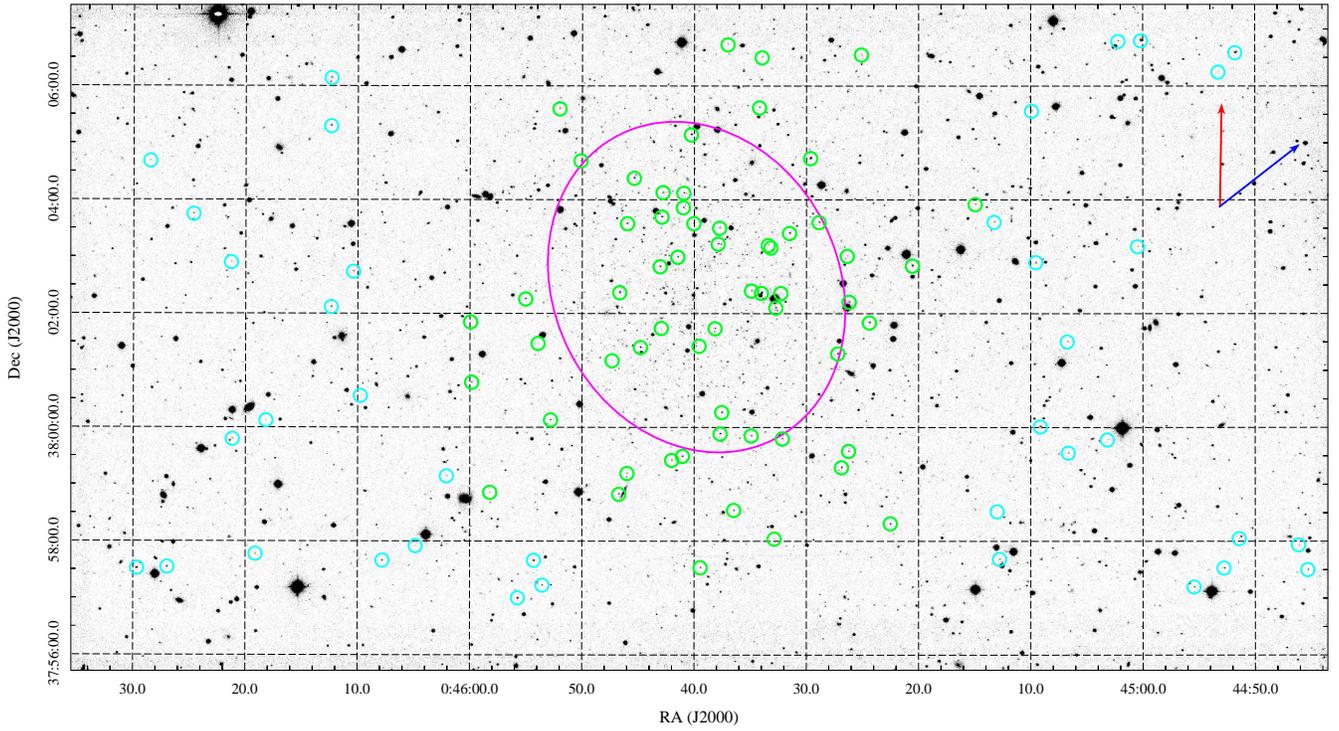}
\caption{The master WFC image of And\,I. The half-light radius is marked with a magenta ellipse. Green circles represent the spatial location of And\,I LPV candidates, and candidates outside two half-light radii from the centre of And\,I are shown as cyan circles. The red and blue arrows are drawn in the direction of the centre and the disc of M\,31 (orthogonal to the major axis), respectively.}
\end{figure*}
%%%%%%%%%%%%%%%%%%%%%%%%%%%%%%%%%%%%%%%%%%%%%%%%%%%%%%%%%%%%%%%%%%%%%%%%%%%%%%%
%@@@@@@@@@@@@@@@@@@@@@@@@@@@@@@@@@@@@@@@@@@@@@@@@@@@@@@@@@@@@@@@ subsection 4.4
%
\subsection{Completeness in LPV candidates catalogues}

%%%%%%%%%%%%%%%%%%%%%%%%%%%%%%%%%%%%%%%%%%%%%%%%%%%%%%%%%%%%%%%%%%%%%%%%%%%%%%%
% FIGURE 18
%
\begin{figure}
\includegraphics[width=1.0\columnwidth]{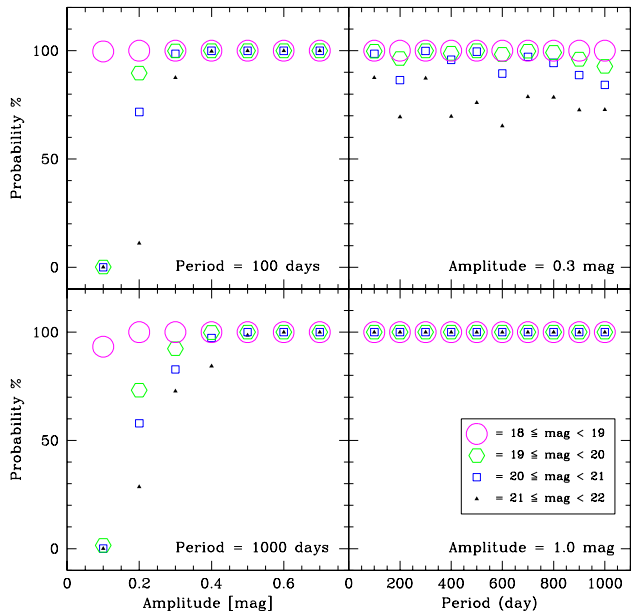}
\caption{Probability of detecting a LPV candidate as a function of amplitude for simulated data with periods of $P=100$ d (top-left panel) and $P=1000$ d (bottom-left panel). The right panels show the probability as a function of period with amplitudes of 0.3 mag (top panel) and 1 mag (bottom panel). Different symbols are used for different magnitude bins.}
\end{figure}
%%%%%%%%%%%%%%%%%%%%%%%%%%%%%%%%%%%%%%%%%%%%%%%%%%%%%%%%%%%%%%%%%%%%%%%%%%%%%%%

In addition to assessing the completeness of the photometric catalogue, which was determined in Sec.\ 3.2, further simulations were used to estimate the completeness of identifying LPVs due to the number of epochs and their timespan. Since the detection probability of a LPV will depend on parameters such as the mean magnitude, period, and amplitude, different sets of simulations were carried out. In each simulation, we created 1000 random sinusoidal light-curves with random first phases. All of these variations had the same amplitude and period, but their mean magnitudes were between 18--22 mag. Then, the variability index $L$ was calculated for magnitudes sampled from these simulated light-curves at similar times as our observations.

We tested the sensitivity of detecting LPVs to the amplitudes, 0.2 to 2 mag at intervals of 0.1 mag for two fixed periods, $P=100$ and 1000 d. The top-left and bottom-left panels of Fig.\ 18 show the detection probability of LPV candidates as a function of amplitude in different magnitude intervals. It is obvious that all of the large-amplitude variables ($>0.5$ mag) are 100\% complete in both periods and most likely for all periods in between. Our LPV candidates catalogues -- except for bright stars -- are incomplete for stars with amplitudes of 0.2 mag or smaller, irrespective of the period.

To estimate the completeness of our catalogues with respect to the period of LPVs we performed our simulations with two fixed amplitudes 0.3 and 1.0 mag and magnitudes in the range between 18--22 mag as before. Periods were chosen at 100 days intervals, ranging from $P=100$ to 1000 d. The measured detection probability as a function of the period is presented in Fig.\ 18. The bottom-right panel of this figure shows all of the large-amplitude variables which will be detected for stars with each period. For smaller amplitude, 0.3 mag, our LPV candidates catalogues are essentially complete as well; in the worst case, for the faintest LPV stars, the completeness limit is 65\% for $P=600$ d.

%@@@@@@@@@@@@@@@@@@@@@@@@@@@@@@@@@@@@@@@@@@@@@@@@@@@@@@@@@@@@@@@@@@@@ section 5
\section{Discussion}

%@@@@@@@@@@@@@@@@@@@@@@@@@@@@@@@@@@@@@@@@@@@@@@@@@@@@@@@@@@@@@@@ subsection 5.1
%
\subsection{Distribution of variable star population}

Our final catalogue contains 9\,824 stars and 97 LPV candidates in the region of CCD 4 of WFC ($11.26\times22.55$ arcmin$^2$) among which 5\,581 stars and 59 LPV candidates are located inside two half-light radii from the centre of And\,I and so are more likely to be members of this dwarf galaxy. Fig.\ 19 shows CMDs of And\,I in the $i$-band and $V$-band {\it vs}.\ $V-i$ colour. Our identified LPV candidates are highlighted in green. There is a clump of LPV candidates between $i\sim20$--22 mag and the tendency of them to the more redder colours is noticeable. Likely dusty AGB stars are highlighted with open circles (cf.\ Sec.\ 4.2).

%%%%%%%%%%%%%%%%%%%%%%%%%%%%%%%%%%%%%%%%%%%%%%%%%%%%%%%%%%%%%%%%%%%%%%%%%%%%%%%
% FIGURE 19
%
\begin{figure}
\includegraphics[width=1.0\columnwidth]{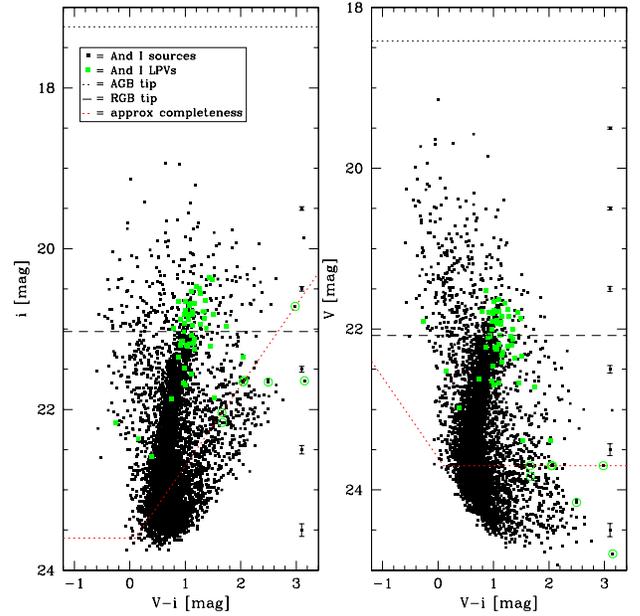}
\caption{CMDs of And\,I in the $i$ and $V$ bands {\it vs}.\ $V-i$ colour showing our identified LPV candidates in green. Likely dusty AGB stars are highlighted with open circles. Average errorbars are plotted for 1 mag intervals in the $i$ and $V$ bands, respectively.}
\end{figure}
%%%%%%%%%%%%%%%%%%%%%%%%%%%%%%%%%%%%%%%%%%%%%%%%%%%%%%%%%%%%%%%%%%%%%%%%%%%%%%%

Various histogram distributions of the identified sources and variables are shown in Fig.\ 20. While the number of stars fainter than $i\sim22$ mag increases, the number of variable stars decreases, so most of them are between $i\sim20$--22 mag; this suggests that many of the fainter stars have not yet evolved to the LPV stage on the AGB. They are post-main sequence stars and likely dominated by low-mass stars which formed many billions of years in the past. At $V-i >1$ mag, LPV candidates are mostly dusty, strongly pulsating, and heavily mass-losing AGB stars.

%%%%%%%%%%%%%%%%%%%%%%%%%%%%%%%%%%%%%%%%%%%%%%%%%%%%%%%%%%%%%%%%%%%%%%%%%%%%%%%
% FIGURE 20
%
\begin{figure}
\includegraphics[width=1.0\columnwidth]{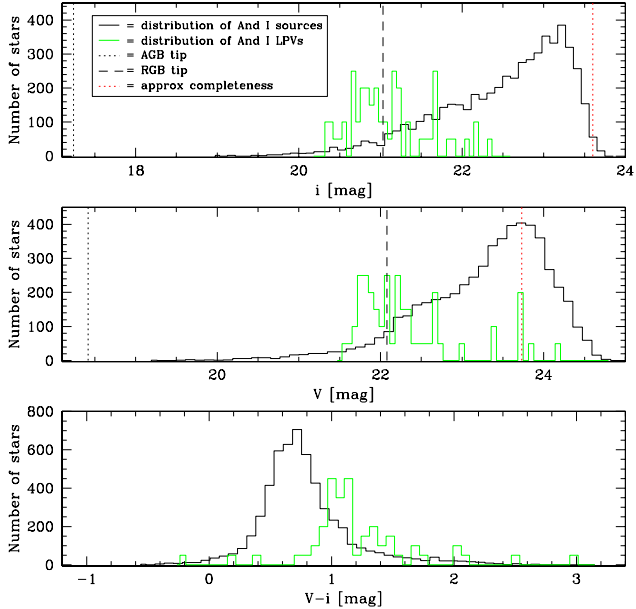}
\caption{Distribution of all And\,I sources (black) and LPV candidates (green) as a function of brightness and colour. The variable stars' histograms are multiplied by 50 for ease of comparison.}
\end{figure}
%%%%%%%%%%%%%%%%%%%%%%%%%%%%%%%%%%%%%%%%%%%%%%%%%%%%%%%%%%%%%%%%%%%%%%%%%%%%%%%
 
Fig.\ 21 presents the identified LPV candidates along with isochrones calculated by Marigo et al.\ (2017). These isochrones are the most appropriate theoretical models for our purpose because they have the most comprehensive treatment of the  thermal pulsating AGB phase, and also predict the variability properties of the stars. The isochrones are calculated for And\,I metallicity, [Fe/H]=$-1.45\pm0.04$ and distance modulus, $\mu=24.41$ mag. We used the reddening value of E(B-V)=0.056 mag (McConnachie et al.\ 2005) to Padova isochrones. The reddening corrections are not applied to our obtained magnitudes in this paper.
We calculated the location of the tip of the RGB at $i=21.03$ mag (cf.\ Sec.\ 5.2). To find the tip of the AGB, we used the classical core--luminosity relation for a Chandrasekhar core mass which is about $M_{\rm bol}=-7.1$ mag. This criterion should not be applied too strictly as AGB stars may exceed this limit, due to the over-luminosity produced during a TP (Zijlstra et al.\ 1996) or as a result of HBB. Based on the Padova models (Marigo et al.\ 2017), this classical AGB limit is equivalent to the peak of the $\sim35$ Myr isochrone which is $i=17.24$ mag for And\,I. Here, the tip of the AGB only serves the purpose of indicating the range where AGB stars may be found; as a dSph, And\,I is not expected to have stars around (or above) the tip of the AGB such as RSGs or AGB stars experiencing HBB. As one can see in Fig.\ 21, most LPV candidates are consistent with isochrones older than 1 Gyr and there are no candidates around 100 Myr. This result lends credibility to our LPV identification method. While Hamren et al.\ (2016) did not find carbon stars in And\,I, we have identified AGB stars around $\sim1$ Gyr that would be expected to be carbon stars.

%%%%%%%%%%%%%%%%%%%%%%%%%%%%%%%%%%%%%%%%%%%%%%%%%%%%%%%%%%%%%%%%%%%%%%%%%%%%%%%
% FIGURE 21
%
\begin{figure}
\includegraphics[width=1.0\columnwidth]{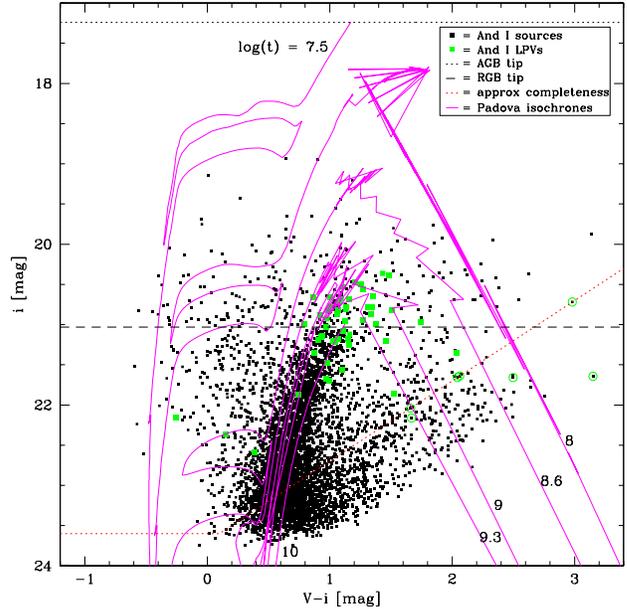}
\caption{CMD of And\,I in the $i$ band {\it vs}.\ $V-i$ colour showing our identified LPV candidates. Overplotted are isochrones from Marigo et al.\ (2017) for a distance modulus of 24.41 mag and metallicity of [Fe/H]=$-1.45$.}
\end{figure}
%%%%%%%%%%%%%%%%%%%%%%%%%%%%%%%%%%%%%%%%%%%%%%%%%%%%%%%%%%%%%%%%%%%%%%%%%%%%%%%

%@@@@@@@@@@@@@@@@@@@@@@@@@@@@@@@@@@@@@@@@@@@@@@@@@@@@@@@@@@@@@@@ subsection 5.2
%
\subsection{Determining the tip of the RGB}

The tip of the RGB is a discontinuity in the RGB luminosity function which can be used as a standard candle. The metal-poor RGB stars in their evolutionary path experience a core helium flash and then leave the RGB branch. The absolute bolometric luminosity of this point is varying only very slightly with mass or metallicity and thus it can be considered to be constant.
The studies have shown that observations in the $I$-band have a minimal dependence on age and chemical composition for determining the tip of the RGB (e.g., Lee, Freedman \& Madore 1993; Salaris \& Cassisi 1998).
To obtain a quantitative estimate of the tip of the RGB, Lee et al.\ (1993) presented a histogram method along with an edge detection filter. Sakai, Madore \& Freedman (1996) used a continuous probability function instead of the histogram to make it independent of binning. Then they convolved this function with a smoothed Sobel kernel in continuous form based on the mean local statistical properties of the data. Here we used a similar method for estimating the tip of the RGB in And\,I. Fig.\ 22 shows a CMD of And\,I in the $I$-band {\it vs}.\ $V-I$ colour (left panel). We converted our data from $i$-band to the $I$-band of Johnson–-Cousins system with transformation equations of Lupton (2005).  
In a similar way as in McConnachie et al.\ (2004), the green dashed lines are determined by eye to select predominantly RGB stars. The binned luminosity function is plotted for the region contained within the dashed lines (with 0.05 mag bins; middle panel). A peak is apparent at I = 20.45 mag in the response of the edge detection filter (right panel) showing the position of the tip of the RGB. 
Adopting a value of $-4.07$ mag for the $I$-band absolute magnitude of the tip of the RGB and Galactic extinction $A_{\rm I}= 1.94E(B-V)$ (McConnachie et al.\ 2005), we obtained a distance modulus, $\mu=24.41\pm0.05$ mag in this study. This result is in good accordance with other studies (cf.\ Sec.\ 1).

%%%%%%%%%%%%%%%%%%%%%%%%%%%%%%%%%%%%%%%%%%%%%%%%%%%%%%%%%%%%%%%%%%%%%%%%%%%%%%%
% FIGURE 22
%
\begin{figure}
\includegraphics[width=1.0\columnwidth]{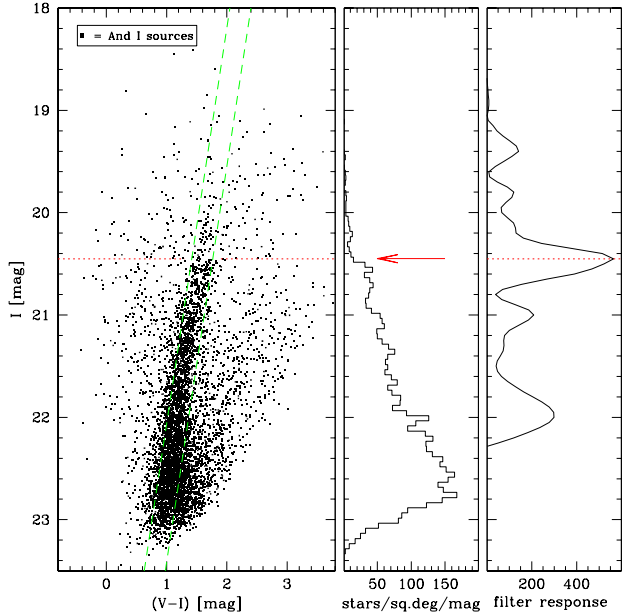}
\caption{Left: CMD of And I sources in the $I$ {\it vs}.\ $V-I$ colour.
Middle: The luminosity function for the region contained within the dashed lines (with 0.05 mag bins). 
Right: response from the edge detection filter. A peak is apparent at I = 20.45 mag.}
\end{figure}
%%%%%%%%%%%%%%%%%%%%%%%%%%%%%%%%%%%%%%%%%%%%%%%%%%%%%%%%%%%%%%%%%%%%%%%%%%%%%%%
%@@@@@@@@@@@@@@@@@@@@@@@@@@@@@@@@@@@@@@@@@@@@@@@@@@@@@@@@@@@@@@@ subsection 5.3
%
\subsection{Cross-correlation with other catalogues}

We cross-correlated our INT variability search results for And\,I with the mid-IR data of the {\it Spitzer} Space Telescope (Boyer et al.\ 2015b) and Wide-field Infrared Survey Explorer (WISE) catalogue (Cutri et al.\ 2013). The matches were obtained in a way similar to that described in Sec.\ 3.1.

%....................................................................... 5.3.1
\subsubsection{{\it Spitzer} mid-IR survey}

Boyer et al.\ (2015a,b) used the {\it Spitzer} Space Telescope to identify dust-producing AGB and massive stars (DUSTiNGS; DUST in Nearby Galaxies with {\it Spitzer}). DUSTiNGS comprised 3.6- and 4.5-$\mu$m imaging of 50 dwarf galaxies within 1.5 Mpc in two epochs, spaced approximately six months apart (Boyer et al.\ 2015a). They classified the identified variables into less dusty AGB and extreme AGB (x-AGB) stars based on their observed IR colour. The x-AGB stars were assumed to be variables brighter than $M_{\rm 3.6}= −8$ mag with colours $[3.6]-[4.5] > 0.1$ mag. These stars make more than 75\% of the dust produced by cool evolved stars, while they include less than 6\% of the total AGB population (e.g., Riebel et al.\ 2012; Boyer et al.\ 2012). In the case of And\,I, DUSTiNGS identified a total of four variable AGB stars, three of which are x-AGB stars (Boyer et al.\ 2015b).

Cross-correlating between our INT photometric catalogue and the DUSTiNGS mid-IR variability survey led to 4\,283 stellar sources in common inside two half-light radii from the centre of And\,I, among which 58 were identified by us as LPV candidates. IR magnitudes exist for about 3/4 of our And\,I optical data and 98\% of LPV candidates. Fig.\ 23 shows a mid-IR CMD of the sources in common, with INT variables highlighted. The red dotted line represents the 75\% completeness limit of the {\it Spitzer} data (Boyer et al.\ 2015a). Isochrones from Marigo et al.\ (2017) for 10 Myr, 100 Myr, 1 Gyr, and 2 Gyr are drawn. LPV candidates are not seen above the tip of the 10 million year isochrone, which is consistent with there being no LPVs among stars more massive than those which become RSGs. All four variable stars from DUSTiNGS are identified in our optical LPV candidates catalogue and presented in asterices on the figure (three x-AGB in red asterices and one less dusty AGB in blue). The blue dashed lines indicate the approximate boundaries between x-AGB stars and others (Boyer et al.\ 2015b). In this work, we found two more, equally x-AGB stars which had not been identified as LPVs in the DUSTiNGS survey.

Another CMD of common sources in the $i$ band {\it vs}.\ $i-[3.6]$ colour (Fig.\ 24) shows five x-AGB stars identified by us. One can discern more candidates of dusty AGB stars with $i-[3.6] \gtrsim 4$ mag on Fig.\ 24 which they have $i$-band amplitudes larger than 1.5 mag. Blum et al.\ (2006) proposed a criterion of $J-[3.6] > 3.1$ mag for x-AGB stars in the Large Magellanic Cloud (LMC). Unfortunately, there is no J-band photometry at the necessary depth for And\,I to use this classification. To get closer to a correct estimate of the number of x-AGB stars, we scaled the LMC down to the size of And\,I in terms of total stellar mass (McConnachie 2012) and number of x-AGB stars (Blum et al.\ 2006). In this way, three x-AGB stars were predicted for And\,I. So we can assume that there will not be many more x-AGB stars in And\,I than those identified by DUSTiNGS and this work, especially given that the star formation rate in the LMC has been higher in the past few hundred million years than that in And\,I.

%%%%%%%%%%%%%%%%%%%%%%%%%%%%%%%%%%%%%%%%%%%%%%%%%%%%%%%%%%%%%%%%%%%%%%%%%%%%%%%
% FIGURE 23
%
\begin{figure}
\includegraphics[width=1.0\columnwidth]{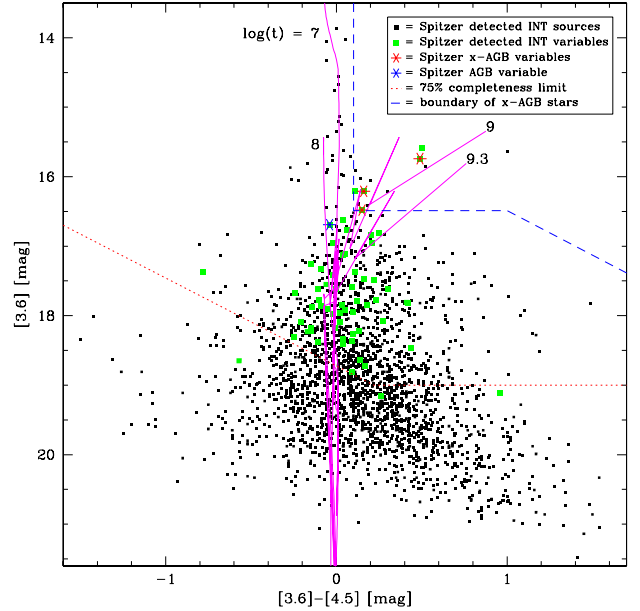}
\caption{Mid-IR CMD from {\it Spitzer}, with INT variables highlighted in green and the variable stars of DUSTiNGS (Boyer et al.\ 2015b) in asterices. Isochrones from Marigo et al.\ (2017) for 10 Myr, 100 Myr, 1 Gyr, and 2 Gyr are drawn. The 75\% completeness limit of these {\it Spitzer} data is shown as a red dotted line (Boyer et al.\ 2015a). The blue dashed lines show the approximate boundaries for x-AGB stars.}
\end{figure}
%%%%%%%%%%%%%%%%%%%%%%%%%%%%%%%%%%%%%%%%%%%%%%%%%%%%%%%%%%%%%%%%%%%%%%%%%%%%%%%
%%%%%%%%%%%%%%%%%%%%%%%%%%%%%%%%%%%%%%%%%%%%%%%%%%%%%%%%%%%%%%%%%%%%%%%%%%%%%%%
% FIGURE 24
%
\begin{figure}
\includegraphics[width=1.0\columnwidth]{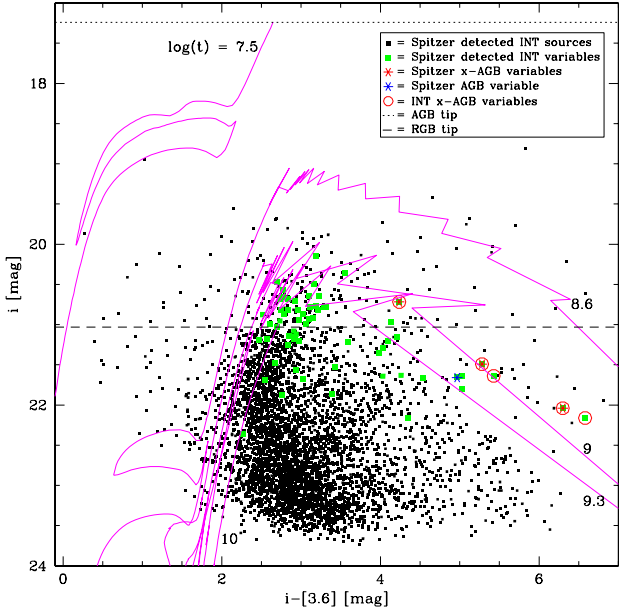}
\caption{CMD of common sources from the DUSTiNGS catalogue and our INT survey in the $i$ band {\it vs}.\ $i-[3.6]$. Five x-AGB stars identified in this work are highlighted with red open circles.}
\end{figure}
%%%%%%%%%%%%%%%%%%%%%%%%%%%%%%%%%%%%%%%%%%%%%%%%%%%%%%%%%%%%%%%%%%%%%%%%%%%%%%%

%....................................................................... 5.3.2
\subsubsection{{\it WISE} mid-IR survey}

WISE is a National Aeronautics and Space Administration (NASA) Medium Class Explorer mission that conducted a digital imaging survey of the entire sky in the $W1\equiv 3.4$, $W2\equiv 4.6$, $W3\equiv 11.6$, and $W4\equiv 22$ $\mu$m mid-IR bandpasses, in 2010 and 2011. The AllWISE program extended the work of the initial WISE mission by combining data from the cryogenic and post-cryogenic survey phases to form the most comprehensive view of the mid-IR sky currently available (Wright et al.\ 2010; Cutri et al.\ 2013).

Within the coverage of CCD\,4 of WFC for And\,I, there are 1\,273 stars from the WISE survey. Out of these, 654 are detected in our INT survey of which 33 were found by us to be variable. Fig.\ 25 shows an optical CMD of our INT survey in which are highlighted with yellow and blue squares the sources in common with {\it Spitzer} data and the WISE survey, respectively. Open red and blue circles represent detected INT variables which were also detected with {\it Spitzer} and WISE. Although the {\it Spitzer} data are better than the WISE data in terms of the number of common sources and photometric quality, the WISE data are still useful because of the W3 and W4 mid-IR bands and the coverage of our data which do not have IR data from {\it Spitzer}.

%%%%%%%%%%%%%%%%%%%%%%%%%%%%%%%%%%%%%%%%%%%%%%%%%%%%%%%%%%%%%%%%%%%%%%%%%%%%%%%
% FIGURE 25
%
\begin{figure}
\includegraphics[width=1.0\columnwidth]{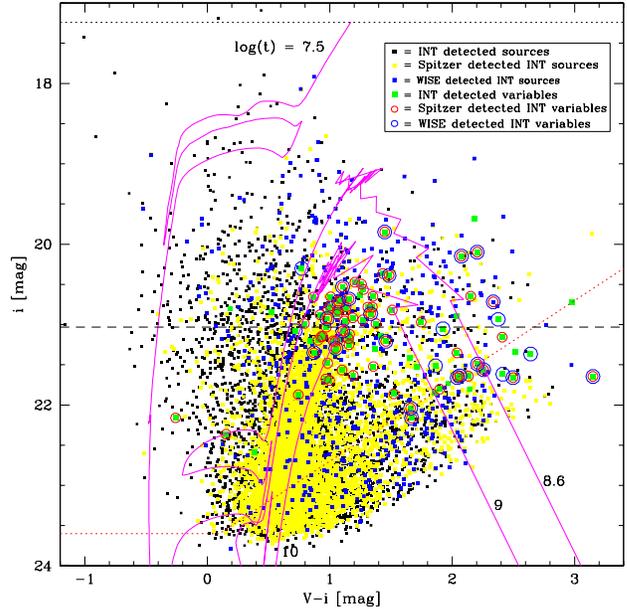}
\caption{Optical CMD showing the stars from our INT survey that were and were not detected in the {\it Spitzer} and WISE surveys.}
\end{figure}
%%%%%%%%%%%%%%%%%%%%%%%%%%%%%%%%%%%%%%%%%%%%%%%%%%%%%%%%%%%%%%%%%%%%%%%%%%%%%%%
%@@@@@@@@@@@@@@@@@@@@@@@@@@@@@@@@@@@@@@@@@@@@@@@@@@@@@@@@@@@@@@@ subsection 5.4
%
\subsection{Estimation of mass-loss rate}

For the purpose of establishing correlations between the mass-loss and dust-production rate on the one hand, and luminosity and amplitude of LPVs on the other, relations between dust optical depth and bolometric corrections, we will model the SEDs of all identified LPV candidates and derive relations for the dust optical depth and bolometric corrections (see Javadi et al.\ 2013). This topic is beyond the scope of the current paper, and we will be dealt with in a future paper in this series. Here, to provide examples of SEDs and an approximate estimate of the mass-return rate of dusty AGB stars, we modelled the SEDs of four dusty LPV candidates from the optical to the IR using the publicly available dust radiative transfer code {\sc dusty} (based on Ivezi\'c \& Elitzur 1997). Fig.\ 26 presents these SEDs which are composed of measurements in the $i$ filter of the INT (these stars had no $V$-band magnitude), four SDSS filters ($u$, $g$, $r$, and $z$), two mid-IR bands of {\it Spitzer} (3.6 and 4.5 $\mu$m), and four mid-IR bands of WISE (W1, W2, W3, W4) with magenta, black, red, and blue points, respectively. Unfortunately, near IR photometry is not available for most of the Andromeda satellites. The horizontal ``errorbars'' on the data represent the width of the photometric passbands. Photometric uncertainties are shown with vertical ``errorbars''.

In the absence of spectroscopic confirmation, we used both types of dust species for matching fits, assuming a grain mixture of 85\% amorphous carbon (Hanner 1988) and 15\% silicon carbide (P\'egouri\'e 1988) for a carbon star (dotted lines) and astronomical silicates (Draine \& Lee 1984) for an O-rich star (solid lines). It is not always possible to determine, based on the fit, which type of star we are dealing with. Different values of optical depth ($\tau$) and luminosity ($L$) were scaled until an acceptable match was obtained, on visual inspection. The SEDs obviously indicate that the W4 and some of the W3 data are too bright in comparison to the other bands. This may have happened due to poor spatial resolution ($6\rlap{.}^{\prime\prime}5$ and $12\rlap{.}^{\prime\prime}0$ for W3 and W4, respectively); we used downward-pointing arrows to show those points which are upper limits. The colours, optical depth, luminosity, and mass-loss rate ($\dot{M}$) of these stars are reported on the figure.

The left panels of Fig.\ 26 show examples of two x-AGB candidates with mass-loss rates of $\sim 2.8\times10^{-6}$ M$_\odot$ yr$^{-1}$. Two examples of likely dusty AGB stars (cf.\ Sec.\ 4.2), however not x-AGB stars, are shown in the right panels. The mass-loss rates of the latter are a little smaller, $\sim 1.6\times10^{-6}$ M$_\odot$ yr$^{-1}$. In a crude approach, we can estimate the mass-return rate from the identified x-AGB stars and other dusty AGB stars. According to the number of these stars in And\,I (five x-AGB and thirteen less dusty AGB stars), the mass-return rate will be $\sim 3.5\times10^{-5}$ M$_\odot$ yr$^{-1}$. In comparison to the total stellar mass of And\,I, this corresponds to $\sim 9\times10^{-12}$ yr$^{-1}$, and so at this rate it would take $\sim 8$ times the age of the Universe to produce this galaxy. Alternatively, we could speculate that And\,I will not grow by more than $\sim10$\% over the coming 10 Gyr or so -- and this is assuming no gas leaves the system.

%%%%%%%%%%%%%%%%%%%%%%%%%%%%%%%%%%%%%%%%%%%%%%%%%%%%%%%%%%%%%%%%%%%%%%%%%%%%%%%
% FIGURE 26
%
\begin{figure}
\vbox{
\includegraphics[width=1.0\columnwidth]{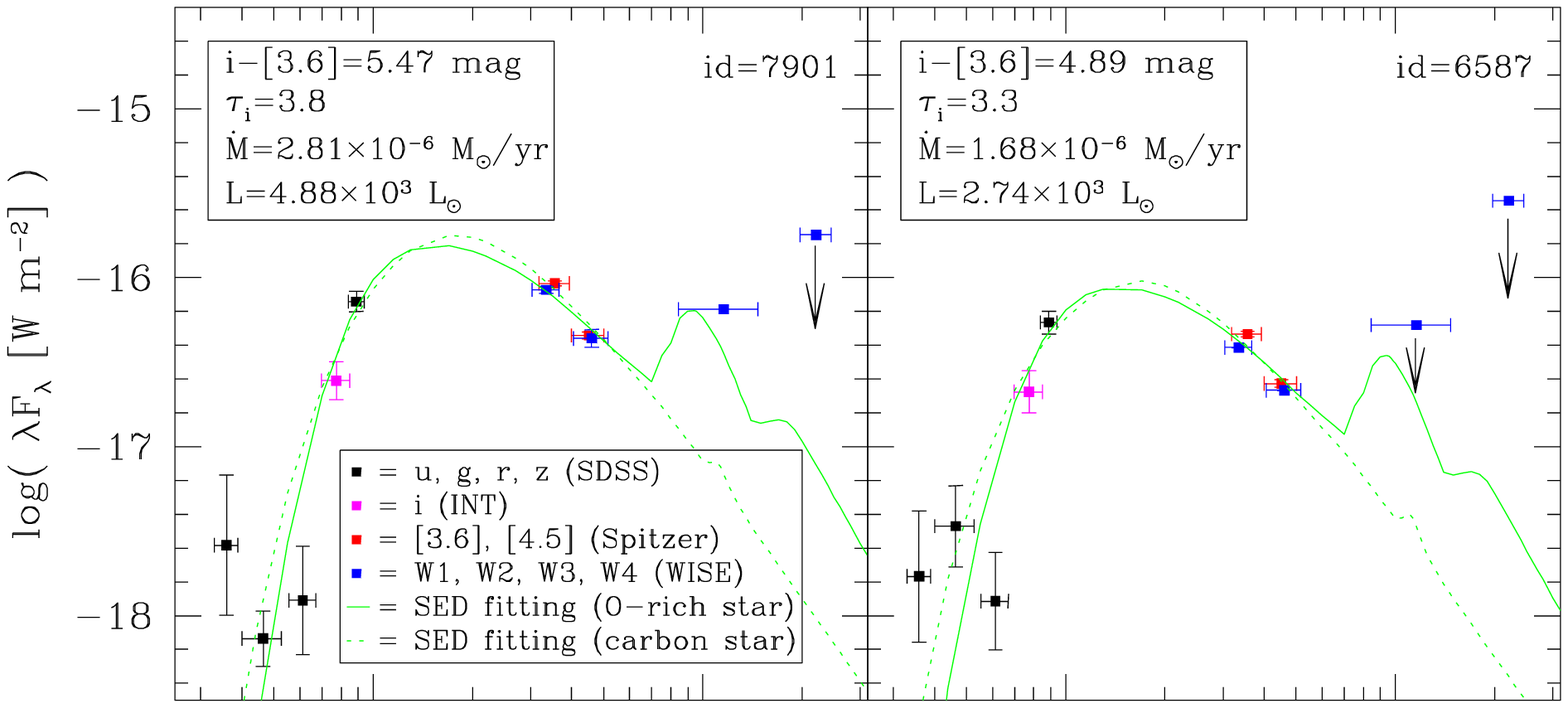}
\includegraphics[width=1.0\columnwidth]{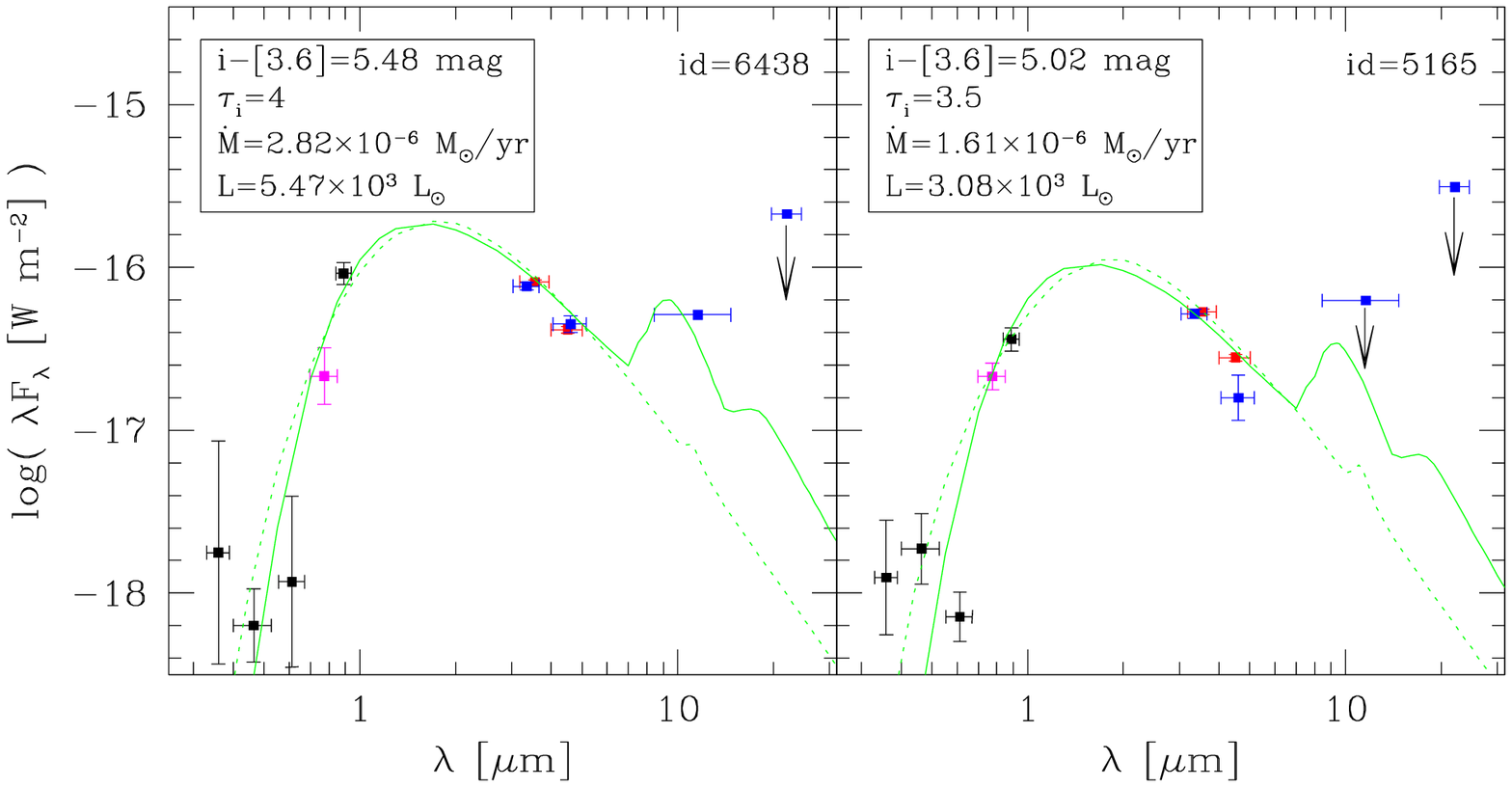}}
\caption{Example SEDs of dusty evolved stars, experiencing various levels of mass loss. The horizontal ``errorbars'' on the data represent the width of the photometric passbands; photometric uncertainties are shown with vertical ``errorbars''. The solid and dotted lines are the best matching {\sc dusty} models assuming an O-rich and a carbon star, respectively.}
\end{figure}
%%%%%%%%%%%%%%%%%%%%%%%%%%%%%%%%%%%%%%%%%%%%%%%%%%%%%%%%%%%%%%%%%%%%%%%%%%%%%%%
%@@@@@@@@@@@@@@@@@@@@@@@@@@@@@@@@@@@@@@@@@@@@@@@@@@@@@@@@@@@@@@@@@@@@ section 6
%
\section{Summary}

The WFC at the INT was used to monitor the majority of dwarf galaxies in the LG including 43 dSph, six dIrr and six dTrans in the $i$-band filter with additional observations in the $V$-band over up to ten epochs. We presented the status of this monitoring survey along with a description of the first results for the And\,I dwarf galaxy to demonstrate the methodology and scientific potential of this project.

A photometric catalogue was constructed containing 9\,824 stars in the region of CCD 4 of WFC ($11.26\times22.55$ arcmin$^2$) with central coordinate $00^{\rm h}45^{\rm m}39\rlap{.}^{\rm s}9$, $+38^\circ02^\prime28^{\prime\prime}$ among which 97 were identified as LPV candidates. The INT catalogue is complete both in photometric data and in terms of finding LPV stars, especially with amplitude more than 0.2 mag. We estimated a distance modulus of $24.41$ mag for And\,I based on the tip of the RGB. Also, a half-light radius of $3.2\pm0.3$ arcmin is calculated. Our catalogue was cross-correlated with the {\it Spitzer} and WISE mid-IR surveys. We identified several dusty AGB stars with red colours and large amplitudes in And\,I, among which five are x-AGB stars. Examples of SED modelling were performed to obtain an estimate of the mass-return rate by these stars.

In the next papers in this series, our catalogue will be used to describe the SFH and dust production for all identified LPV candidates in the entire sample of monitored galaxies. Also, we will investigate how the colour and hence temperature changes during the variability; from temperature and luminosity, we will determine how the radius varies, and how that is related to the mid-IR excess.

%==============================================================================
\section*{Acknowledgments}

The observing time for this survey was primarily provided by the Iranian National Observatory (INO), complemented by UK-PATT allocation of time to programmes I/2016B/09 and I/2017B/04 (PI: J.\ van Loon). We thank the INO and the School of Astronomy (IPM) for the financial support of this project. We thank the referee for their comments which helped enhance the analysis. ES is grateful to Peter Stetson for sharing his photometry routines. JvL thanks the IPM for their hospitality during his visits, some of which were funded by the Royal Society International Exchange, grant IE130487; he also acknowledges a Santander Travel Bursary and a PATT grant to travel to the Canary Islands. JB acknowledges an STFC studentship at Keele University. IMcD is supported by the United Kingdom's Science and Technology Facilities Council (STFC), grant ST/P000649/1.

%==============================================================================

%@@@@@@@@@@@@@@@@@@@@@@@@@@@@@@@@@@@@@@@@@@@@@@@@@@@@@@@@@@@@@@@@@@@@@ appendix
\appendix
%--------------------------------------------------------------------------- A1
\section{Supplementary material}
\label{app:app}

%
% TABLE 2
%
\begin{table*}
\centering
\caption{Log of WFC observations of each of 60 fields.}
\begin{tabular}{lcccccc|lcccccc}
	\hline\hline
 \noalign{\smallskip}
	Galaxy                            &
	Date                              &
	E                                 &
	F                                 &
	$t_{\rm exp}$                     &
        Airmass                           &
        seeing                            &
	Galaxy                            &
	Date                              &
	E                                 &
	F                                 &
	$t_{\rm exp}$                     &
        Airmass                           &
        seeing                            \\
        &
        (y\,m\,d)                &
        &
        &
        (sec)                    &
        &
        (arcsec)
        &
        &               
        (y\,m\,d)                &
        &
        &
        (sec)                    &
        &
        (arcsec)                \\ 
 \noalign{\smallskip}
	\hline
\noalign{\smallskip}
And\,I      &  2016 02 09  &  2  &  $i$  &  540   &  1.475  &  1.35   &  And\,VII    &  2017 09 02  &  8  &  $i$  &  555   &  1.089  &  1.21  \\
And\,I      &  2016 06 14  &  3  &  $i$  &  555   &  1.568  &  1.68   &  And\,VII    &  2017 09 02  &  8  &  $V$  &  735   &  1.080  &  1.31  \\
And\,I      &  2016 08 10  &  4  &  $i$  &  555   &  1.077  &  1.39   &  And\,VII    &  2017 10 06  &  9  &  $i$  &  555   &  1.185  &  1.34  \\
And\,I      &  2016 08 12  &  4  &  $V$  &  735   &  1.015  &  1.30   &  And\,VII    &  2017 10 08  &  9  &  $V$  &  735   &  1.196  &  1.66  \\
And\,I      &  2016 10 20  &  5  &  $i$  &  555   &  1.205  &  1.48   &  And\,IX     &  2015 06 21  &  1  &  $I$  &  45    &  1.265  &  1.65  \\
And\,I      &  2016 10 20  &  5  &  $V$  &  735   &  1.286  &  1.53   &  And\,IX     &  2015 06 21  &  1  &  $V$  &  72    &  1.253  &  1.76  \\
And\,I      &  2017 01 29  &  6  &  $i$  &  555   &  2.425  &  1.77   &  And\,IX     &  2016 02 09  &  2  &  $i$  &  540   &  1.854  &  1.32  \\
And\,I      &  2017 08 01  &  7  &  $i$  &  555   &  1.018  &  1.18   &  And\,IX     &  2016 06 15  &  3  &  $i$  &  555   &  1.409  &  1.76  \\  
And\,I      &  2017 08 01  &  7  &  $V$  &  735   &  1.014  &  1.25   &  And\,IX     &  2016 08 11  &  4  &  $i$  &  555   &  1.045  &  1.20  \\  
And\,I      &  2017 09 01  &  8  &  $i$  &  555   &  1.084  &  1.25   &  And\,IX     &  2016 08 13  &  4  &  $V$  &  735   &  1.033  &  1.50  \\   
And\,I      &  2017 09 01  &  8  &  $V$  &  735   &  1.051  &  1.34   &  And\,IX     &  2016 10 21  &  5  &  $i$  &  555   &  1.320  &  1.00  \\   
And\,I      &  2017 10 06  &  9  &  $i$  &  555   &  1.278  &  1.44   &  And\,IX     &  2017 01 29  &  6  &  $i$  &  435   &  1.655  &  1.53  \\  
And\,I      &  2017 10 08  &  9  &  $V$  &  735   &  1.014  &  1.21   &  And\,IX     &  2017 08 01  &  7  &  $i$  &  555   &  1.306  &  1.22  \\   

And\,II     &  2016 02 09  &  2  &  $i$  &  540   &  1.968  &  1.28   &  And\,IX     &  2017 08 01  &  7  &  $V$  &  735   &  1.227  &  1.48  \\  
And\,II     &  2016 06 13  &  3  &  $i$  &  565   &  2.550  &  1.78   &  And\,IX     &  2017 09 02  &  8  &  $i$  &  555   &  1.070  &  1.23  \\ 
And\,II     &  2016 08 12  &  4  &  $i$  &  555   &  1.335  &  1.42   &  And\,IX     &  2017 09 02  &  8  &  $V$  &  735   &  1.047  &  1.24  \\  
And\,II     &  2016 08 13  &  4  &  $V$  &  735   &  1.052  &  1.70   &  And\,IX     &  2017 10 06  &  9  &  $i$  &  555   &  1.341  &  1.04  \\    
And\,II     &  2016 10 20  &  5  &  $i$  &  555   &  1.522  &  1.28   &  And\,X      &  2016 02 08  &  2  &  $i$  &  810   &  1.827  &  1.77  \\  
And\,II     &  2016 10 22  &  5  &  $V$  &  735   &  1.405  &  1.34   &  And\,X      &  2016 06 13  &  3  &  $i$  &  615   &  2.488  &  2.68  \\ 
And\,II     &  2017 01 29  &  6  &  $i$  &  495   &  1.723  &  1.61   &  And\,X      &  2016 08 10  &  4  &  $i$  &  555   &  1.052  &  1.17  \\  
And\,II     &  2017 08 02  &  7  &  $i$  &  555   &  1.091  &  1.16   &  And\,X      &  2016 08 12  &  4  &  $V$  &  735   &  1.102  &  1.45  \\  
And\,II     &  2017 09 01  &  8  &  $i$  &  555   &  1.046  &  1.40   &  And\,X      &  2016 10 21  &  5  &  $i$  &  555   &  1.368  &  0.90  \\  
And\,II     &  2017 09 01  &  8  &  $V$  &  735   &  1.022  &  1.39   &  And\,X      &  2016 10 22  &  5  &  $V$  &  735   &  1.238  &  1.20  \\  
And\,II     &  2017 10 06  &  9  &  $i$  &  555   &  1.061  &  1.00   &  And\,X      &  2017 01 29  &  6  &  $i$  &  555   &  1.424  &  1.46  \\  
And\,II     &  2017 10 08  &  9  &  $V$  &  735   &  1.402  &  1.29   &  And\,X      &  2017 08 01  &  7  &  $i$  &  555   &  1.104  &  1.20  \\

And\,III    &  2016 02 09  &  2  &  $i$  &  540   &  1.676  &  1.39   &  And\,X      &  2017 08 01  &  7  &  $V$  &  735   &  1.074  &  1.29  \\  
And\,III    &  2016 06 15  &  3  &  $i$  &  555   &  1.455  &  1.69   &  And\,X      &  2017 09 02  &  8  &  $i$  &  555   &  1.049  &  1.14  \\  
And\,III    &  2016 08 10  &  4  &  $i$  &  555   &  1.020  &  1.22   &  And\,X      &  2017 09 02  &  8  &  $V$  &  735   &  1.041  &  1.19  \\  
And\,III    &  2016 08 12  &  4  &  $V$  &  735   &  1.034  &  1.51   &  And\,X      &  2017 10 06  &  9  &  $i$  &  555   &  1.148  &  0.87  \\  
And\,III    &  2016 10 20  &  5  &  $i$  &  555   &  1.166  &  1.10   &  And\,X      &  2017 10 09  &  9  &  $V$  &  735   &  1.159  &  1.09  \\   
And\,III    &  2016 10 22  &  5  &  $V$  &  735   &  1.061  &  0.95   &  And\,XI     &  2016 08 09  &  4  &  $i$  &  555   &  1.024  &  2.61  \\  
And\,III    &  2017 01 30  &  6  &  $i$  &  495   &  1.650  &  1.81   &  And\,XI     &  2016 08 09  &  4  &  $V$  &  735   &  1.008  &  2.84  \\ 
And\,III    &  2017 01 31  &  6  &  $i$  &  540   &  2.281  &  2.76   &  And\,XI     &  2016 08 11  &  4  &  $i$  &  555   &  1.032  &  1.24  \\  
And\,III    &  2017 08 02  &  7  &  $i$  &  555   &  1.109  &  1.14   &  And\,XI     &  2016 08 13  &  4  &  $V$  &  735   &  1.053  &  1.76  \\  
And\,III    &  2017 09 01  &  8  &  $i$  &  975   &  1.095  &  1.44   &  And\,XI     &  2016 10 19  &  5  &  $i$  &  555   &  1.295  &  1.06  \\  
And\,III    &  2017 09 01  &  8  &  $V$  &  735   &  1.165  &  1.41   &  And\,XI     &  2016 10 22  &  5  &  $V$  &  735   &  1.081  &  1.05  \\  
And\,III    &  2017 10 06  &  9  &  $i$  &  555   &  1.013  &  1.16   &  And\,XI     &  2017 01 30  &  6  &  $i$  &  555   &  2.556  &  2.41  \\  
And\,III    &  2017 10 09  &  9  &  $V$  &  735   &  1.485  &  1.55   &  And\,XI     &  2017 08 01  &  7  &  $i$  &  555   &  1.133  &  1.44  \\

And\,V      &  2016 02 08  &  2  &  $i$  &  540   &  1.572  &  1.60   &  And\,XI     &  2017 08 01  &  7  &  $V$  &  735   &  1.083  &  1.43  \\  
And\,V      &  2016 06 15  &  3  &  $i$  &  555   &  1.386  &  1.56   &  And\,XI     &  2017 09 02  &  8  &  $i$  &  555   &  1.014  &  1.03  \\   
And\,V      &  2016 08 12  &  4  &  $i$  &  555   &  1.168  &  1.18   &  And\,XI     &  2017 09 02  &  8  &  $V$  &  735   &  1.033  &  1.17  \\  
And\,V      &  2016 08 13  &  4  &  $V$  &  555   &  1.071  &  1.49   &  And\,XI     &  2017 10 06  &  9  &  $i$  &  555   &  1.029  &  1.09  \\  
And\,V      &  2016 10 21  &  5  &  $i$  &  300   &  1.433  &  1.22   &  And\,XI     &  2017 10 08  &  9  &  $V$  &  735   &  1.282  &  1.11  \\  
And\,V      &  2016 10 22  &  5  &  $V$  &  735   &  1.319  &  1.11   &  And\,XII    &  2016 02 10  &  2  &  $i$  &  630   &  2.159  &  1.67  \\  
And\,V      &  2017 01 29  &  6  &  $i$  &  555   &  2.558  &  1.86   &  And\,XII    &  2016 06 14  &  3  &  $i$  &  480   &  1.283  &  1.83  \\  
And\,V      &  2017 08 03  &  7  &  $i$  &  555   &  1.060  &  1.48   &  And\,XII    &  2016 08 11  &  4  &  $i$  &  555   &  1.062  &  1.26  \\    
And\,V      &  2017 08 03  &  7  &  $V$  &  735   &  1.057  &  1.40   &  And\,XII    &  2016 08 13  &  4  &  $V$  &  735   &  1.097  &  1.85  \\  
And\,V      &  2017 09 02  &  8  &  $i$  &  555   &  1.614  &  1.32   &  And\,XII    &  2016 10 21  &  5  &  $i$  &  555   &  1.068  &  0.93  \\  
And\,V      &  2017 09 02  &  8  &  $V$  &  735   &  1.477  &  1.56   &  And\,XII    &  2016 10 21  &  5  &  $V$  &  735   &  1.110  &  1.03  \\  
And\,V      &  2017 10 06  &  9  &  $i$  &  555   &  1.214  &  0.99   &  And\,XII    &  2017 01 31  &  6  &  $i$  &  555   &  1.664  &  2.00  \\  
And\,V      &  2017 10 09  &  9  &  $V$  &  735   &  1.311  &  1.34   &  And\,XII    &  2017 08 03  &  7  &  $i$  &  555   &  1.021  &  1.47  \\
  
And\,VI     &  2016 06 13  &  3  &  $i$  &  555   &  2.270  &  2.22   &  And\,XII    &  2017 09 02  &  8  &  $i$  &  555   &  1.063  &  1.01  \\ 
And\,VI     &  2016 08 10  &  4  &  $i$  &  555   &  1.132  &  1.35   &  And\,XII    &  2017 09 02  &  8  &  $V$  &  735   &  1.104  &  1.10  \\  
And\,VI     &  2016 08 13  &  4  &  $V$  &  735   &  1.356  &  1.74   &  And\,XII    &  2017 10 06  &  9  &  $i$  &  555   &  1.038  &  0.99  \\  
And\,VI     &  2016 10 20  &  5  &  $i$  &  555   &  1.003  &  1.30   &  And\,XII    &  2017 10 09  &  9  &  $V$  &  735   &  1.276  &  1.36  \\   
And\,VI     &  2016 10 22  &  5  &  $V$  &  735   &  1.015  &  1.26   &  And\,XIII   &  2016 02 09  &  2  &  $i$  &  540   &  1.759  &  1.43  \\  
And\,VI     &  2017 01 30  &  6  &  $i$  &  480   &  1.731  &  1.55   &  And\,XIII   &  2016 06 14  &  3  &  $i$  &  555   &  1.478  &  1.47  \\  
And\,VI     &  2017 02 01  &  6  &  $i$  &  540   &  1.701  &  1.33   &  And\,XIII   &  2016 08 11  &  4  &  $i$  &  555   &  1.134  &  1.29  \\  
And\,VI     &  2017 08 02  &  7  &  $i$  &  555   &  1.408  &  1.34   &  And\,XIII   &  2016 08 13  &  4  &  $V$  &  735   &  1.235  &  1.80  \\  
And\,VI     &  2017 08 02  &  7  &  $V$  &  735   &  1.289  &  1.40   &  And\,XIII   &  2016 10 21  &  5  &  $i$  &  555   &  1.159  &  0.99  \\   
And\,VI     &  2017 09 02  &  8  &  $i$  &  555   &  1.129  &  1.34   &  And\,XIII   &  2016 10 21  &  5  &  $V$  &  735   &  1.233  &  1.03  \\  
And\,VI     &  2017 09 02  &  8  &  $V$  &  735   &  1.072  &  1.32   &  And\,XIII   &  2017 01 31  &  6  &  $i$  &  555   &  1.473  &  1.75  \\  
And\,VI     &  2017 10 06  &  9  &  $i$  &  555   &  1.080  &  1.10   &  And\,XIII   &  2017 08 02  &  7  &  $i$  &  555   &  1.012  &  1.00  \\  
And\,VI     &  2017 10 09  &  9  &  $V$  &  735   &  1.471  &  1.83   &  And\,XIII   &  2017 09 02  &  8  &  $i$  &  555   &  1.148  &  1.00  \\

And\,VII    &  2015 06 18  &  1  &  $I$  &  540   &  1.328  &  1.21   &  And\,XIII   &  2017 09 02  &  8  &  $V$  &  735   &  1.217  &  1.11  \\  
And\,VII    &  2015 06 18  &  1  &  $V$  &  1200  &  1.174  &  1.29   &  And\,XIII   &  2017 10 06  &  9  &  $i$  &  555   &  1.004  &  1.27  \\  
And\,VII    &  2016 02 10  &  2  &  $i$  &  630   &  1.894  &  1.45   &  And\,XIII   &  2017 10 09  &  9  &  $V$  &  735   &  1.433  &  1.40  \\   
And\,VII    &  2016 06 13  &  3  &  $i$  &  615   &  2.214  &  2.84   &  And\,XIV    &  2016 08 11  &  4  &  $i$  &  555   &  1.000  &  1.15  \\  
And\,VII    &  2016 08 10  &  4  &  $i$  &  555   &  1.198  &  1.54   &  And\,XIV    &  2016 08 13  &  4  &  $V$  &  735   &  1.006  &  1.64  \\  
And\,VII    &  2016 08 12  &  4  &  $i$  &  555   &  1.287  &  1.37   &  And\,XIV    &  2016 10 19  &  5  &  $i$  &  555   &  1.394  &  1.09  \\  
And\,VII    &  2016 08 12  &  4  &  $V$  &  735   &  1.076  &  1.24   &  And\,XIV    &  2016 10 22  &  5  &  $V$  &  735   &  1.123  &  1.01  \\  
And\,VII    &  2016 10 20  &  5  &  $i$  &  555   &  1.103  &  1.55   &  And\,XIV    &  2017 01 31  &  6  &  $i$  &  555   &  1.268  &  1.65  \\  
And\,VII    &  2017 01 30  &  6  &  $i$  &  495   &  1.581  &  1.23   &  And\,XIV    &  2017 08 02  &  7  &  $i$  &  555   &  1.001  &  1.09  \\  
And\,VII    &  2017 08 01  &  7  &  $i$  &  555   &  1.724  &  1.23   &  And\,XIV    &  2017 09 03  &  8  &  $i$  &  555   &  1.439  &  1.25  \\   
And\,VII    &  2017 08 01  &  7  &  $V$  &  735   &  1.571  &  1.44   &  And\,XIV    &  2017 09 03  &  8  &  $V$  &  735   &  1.310  &  1.43  \\
\noalign{\smallskip}
\hline
\end{tabular}
\end{table*}

\begin{table*}
\centering
	%\contcaption{}
\begin{tabular}{lcccccc|lcccccc}
	\hline\hline
 \noalign{\smallskip}
	Galaxy                            &
	Date                              &
	E                                 &
	F                                 &
	$t_{\rm exp}$                     &
        Airmass                           &
        seeing                            &
	Galaxy                            &
	Date                              &
	E                                 &
	F                                 &
	$t_{\rm exp}$                     &
        Airmass                           &
        seeing                            \\
        &
        (y\,m\,d)                &
        &
        &
        (sec)                    &
        &
        (arcsec)
        &
        &               
        (y\,m\,d)                &
        &
        &
        (sec)                    &
        &
        (arcsec)                \\ 
 \noalign{\smallskip}
	\hline
\noalign{\smallskip}
And\,XIV    &  2017 10 07  &  9  &  $i$  &  555   &  1.006  &  1.80   &  And\,XXI    &  2017 01 31  &  6  &  $i$  &  540   &  2.443  &  2.05  \\ 
And\,XIV    &  2017 10 08  &  9  &  $V$  &  735   &  1.404  &  1.15   &  And\,XXI    &  2017 08 01  &  7  &  $i$  &  555   &  1.590  &  1.49  \\  

And\,XV     &  2016 02 08  &  2  &  $i$  &  540   &  1.455  &  1.59   &  And\,XXI    &  2017 08 01  &  7  &  $V$  &  735   &  1.446  &  1.79  \\
And\,XV     &  2016 08 11  &  4  &  $i$  &  555   &  1.021  &  1.09   &  And\,XXI    &  2017 09 03  &  8  &  $i$  &  555   &  1.141  &  0.97  \\
And\,XV     &  2016 08 13  &  4  &  $V$  &  735   &  1.021  &  1.54   &  And\,XXI    &  2017 09 03  &  8  &  $V$  &  735   &  1.201  &  1.14  \\
And\,XV     &  2016 10 19  &  5  &  $i$  &  555   &  1.502  &  1.07   &  And\,XXI    &  2017 10 07  &  9  &  $i$  &  555   &  1.060  &  1.73  \\
And\,XV     &  2017 01 29  &  6  &  $i$  &  555   &  1.288  &  1.43   &  And\,XXI    &  2017 10 08  &  9  &  $V$  &  735   &  1.125  &  0.99  \\
And\,XV     &  2017 08 02  &  7  &  $i$  &  555   &  1.061  &  1.06   &  And\,XXII   &  2017 01 29  &  6  &  $i$  &  555   &  2.422  &  1.68  \\
And\,XV     &  2017 09 03  &  8  &  $i$  &  555   &  1.285  &  1.07   &  And\,XXII   &  2017 08 02  &  7  &  $i$  &  555   &  1.174  &  1.18  \\
And\,XV     &  2017 09 03  &  8  &  $V$  &  735   &  1.187  &  1.30   &  And\,XXII   &  2017 09 03  &  8  &  $i$  &  555   &  1.059  &  1.00  \\ 
And\,XV     &  2017 10 06  &  9  &  $i$  &  555   &  1.044  &  1.05   &  And\,XXII   &  2017 09 03  &  8  &  $V$  &  735   &  1.104  &  1.16  \\
And\,XV     &  2017 10 08  &  9  &  $V$  &  735   &  1.128  &  1.17   &  And\,XXII   &  2017 10 07  &  9  &  $i$  &  555   &  1.070  &  1.57  \\

And\,XVI    &  2016 02 10  &  2  &  $i$  &  540   &  2.322  &  1.64   &  M\,32       &  2016 06 14  &  3  &  $i$  &  555   &  1.703  &  1.68  \\
And\,XVI    &  2016 08 12  &  4  &  $i$  &  465   &  1.174  &  1.12   &  M\,32       &  2016 10 21  &  5  &  $i$  &  555   &  1.065  &  1.03  \\
And\,XVI    &  2016 10 19  &  5  &  $i$  &  465   &  1.472  &  1.09   &  M\,32       &  2016 10 21  &  5  &  $V$  &  735   &  1.042  &  1.32  \\
And\,XVI    &  2016 10 22  &  5  &  $V$  &  735   &  1.168  &  1.07   &  M\,32       &  2017 01 31  &  6  &  $i$  &  555   &  2.416  &  2.80  \\
And\,XVI    &  2017 01 31  &  6  &  $i$  &  465   &  1.322  &  1.56   &  M\,32       &  2017 10 07  &  9  &  $i$  &  555   &  1.358  &  1.76  \\
And\,XVI    &  2017 08 03  &  7  &  $i$  &  465   &  1.012  &  1.55   &  M\,110      &  2016 10 21  &  5  &  $i$  &  465   &  1.031  &  1.13  \\
And\,XVI    &  2017 09 03  &  8  &  $i$  &  555   &  1.082  &  1.07   &  M\,110      &  2016 10 21  &  5  &  $V$  &  915   &  1.026  &  1.19  \\ 
And\,XVI    &  2017 09 03  &  8  &  $V$  &  735   &  1.041  &  1.28   &  M\,110      &  2017 01 31  &  6  &  $i$  &  180   &  2.792  &  2.50  \\
And\,XVI    &  2017 10 06  &  9  &  $i$  &  465   &  1.074  &  1.18   &  M\,110      &  2017 10 07  &  9  &  $i$  &  555   &  1.251  &  1.70  \\ 
And\,XVI    &  2017 10 08  &  9  &  $V$  &  735   &  1.082  &  1.17   &  IC\,10      &  2015 06 21  &  1  &  $I$  &  430   &  1.748  &  1.87  \\

And\,XVII   &  2015 06 21  &  1  &  $I$  &  104   &  1.312  &  1.57   &  IC\,10      &  2015 06 21  &  1  &  $V$  &  430   &  1.634  &  1.90  \\
And\,XVII   &  2015 06 21  &  1  &  $V$  &  126   &  1.259  &  1.60   &  IC\,10      &  2016 06 13  &  3  &  $i$  &  555   &  1.450  &  1.10  \\
And\,XVII   &  2016 02 10  &  2  &  $i$  &  630   &  1.889  &  1.69   &  IC\,10      &  2016 06 15  &  3  &  $i$  &  555   &  2.271  &  1.82  \\ 
And\,XVII   &  2016 06 15  &  3  &  $i$  &  555   &  1.591  &  1.70   &  IC\,10      &  2016 08 10  &  4  &  $i$  &  555   &  1.168  &  1.11  \\ 
And\,XVII   &  2016 08 11  &  4  &  $i$  &  555   &  1.226  &  1.26   &  IC\,10      &  2016 08 13  &  4  &  $V$  &  735   &  1.331  &  1.69  \\
And\,XVII   &  2016 08 13  &  4  &  $V$  &  735   &  1.385  &  1.66   &  IC\,10      &  2016 08 13  &  4  &  $V$  &  735   &  1.331  &  1.69  \\
And\,XVII   &  2016 10 21  &  5  &  $i$  &  555   &  1.078  &  0.98   &  IC\,10      &  2016 10 19  &  5  &  $i$  &  555   &  1.303  &  1.55  \\
And\,XVII   &  2017 01 30  &  6  &  $i$  &  360   &  2.760  &  2.55   &  IC\,10      &  2016 10 19  &  5  &  $V$  &  735   &  1.361  &  1.62  \\
And\,XVII   &  2017 08 03  &  7  &  $i$  &  555   &  1.069  &  1.47   &  IC\,10      &  2017 08 01  &  7  &  $i$  &  555   &  1.496  &  1.35  \\
And\,XVII   &  2017 09 03  &  8  &  $i$  &  555   &  1.596  &  1.32   &  IC\,10      &  2017 08 01  &  7  &  $V$  &  735   &  1.411  &  1.40  \\
And\,XVII   &  2017 09 03  &  8  &  $V$  &  735   &  1.456  &  1.39   &  IC\,10      &  2017 09 01  &  8  &  $i$  &  555   &  1.263  &  1.47  \\
And\,XVII   &  2017 10 07  &  9  &  $i$  &  555   &  1.049  &  1.27   &  IC\,10      &  2017 09 01  &  8  &  $V$  &  735   &  1.225  &  1.62  \\

And\,XVIII  &  2015 06 21  &  1  &  $I$  &  269   &  1.343  &  1.67   &  IC\,10      &  2017 09 02  &  8  &  $i$  &  555   &  1.634  &  1.18  \\
And\,XVIII  &  2015 06 21  &  1  &  $V$  &  195   &  1.279  &  1.81   &  IC\,10      &  2017 09 02  &  8  &  $V$  &  735   &  1.525  &  1.43  \\
And\,XVIII  &  2016 06 14  &  3  &  $i$  &  2730  &  2.117  &  1.62   &  IC\,10      &  2017 10 06  &  9  &  $i$  &  555   &  1.263  &  1.24  \\
And\,XVIII  &  2016 10 20  &  5  &  $i$  &  2715  &  1.128  &  1.40   &  IC\,10      &  2017 10 08  &  9  &  $V$  &  735   &  1.184  &  1.11  \\
And\,XVIII  &  2017 08 03  &  7  &  $i$  &  2730  &  1.285  &  1.55   &  NGC\,2419   &  2016 02 07  &  2  &  $i$  &  45    &  1.171  &  1.09  \\
And\,XVIII  &  2017 08 03  &  7  &  $V$  &  3180  &  1.085  &  1.76   &  NGC\,2419   &  2016 02 07  &  2  &  $V$  &  45    &  1.145  &  1.15  \\
And\,XVIII  &  2017 10 07  &  9  &  $i$  &  900   &  1.157  &  1.61   &  NGC\,2419   &  2016 02 08  &  2  &  $i$  &  180   &  1.061  &  1.43  \\
And\,XVIII  &  2017 10 08  &  9  &  $i$  &  2730  &  1.053  &  1.43   &  NGC\,2419   &  2016 10 20  &  5  &  $i$  &  45    &  1.143  &  1.51  \\

And\,XIX    &  2016 02 10  &  2  &  $i$  &  630   &  1.962  &  1.70   &  NGC\,2419   &  2016 10 20  &  5  &  $V$  &  45    &  1.117  &  1.57  \\
And\,XIX    &  2016 06 14  &  3  &  $i$  &  555   &  1.780  &  1.72   &  NGC\,2419   &  2017 01 29  &  6  &  $i$  &  45    &  1.016  &  1.19  \\
And\,XIX    &  2016 08 11  &  4  &  $i$  &  555   &  1.032  &  1.13   &  NGC\,2419   &  2017 01 31  &  6  &  $V$  &  60    &  1.016  &  1.55  \\
And\,XIX    &  2016 08 13  &  4  &  $V$  &  320   &  1.089  &  1.50   &  NGC\,2419   &  2017 10 06  &  9  &  $i$  &  45    &  1.185  &  1.41  \\
And\,XIX    &  2016 10 20  &  5  &  $i$  &  555   &  1.145  &  1.13   &  NGC\,2419   &  2017 10 09  &  9  &  $V$  &  45    &  1.108  &  1.25  \\
And\,XIX    &  2016 10 22  &  5  &  $V$  &  735   &  1.039  &  1.04   &  NGC\,2419   &  2018 02 24  &  10 &  $i$  &  45    &  1.032  &  1.79  \\
And\,XIX    &  2017 01 30  &  6  &  $i$  &  555   &  2.536  &  1.77   &  NGC\,2419   &  2018 02 24  &  10 &  $V$  &  45    &  1.041  &  2.44  \\
And\,XIX    &  2017 08 02  &  7  &  $i$  &  555   &  1.126  &  1.21   &  Cetus       &  2016 10 21  &  5  &  $i$  &  540   &  1.323  &  1.00  \\  
And\,XIX    &  2017 09 03  &  8  &  $i$  &  555   &  1.006  &  1.20   &  Cetus       &  2016 10 21  &  5  &  $V$  &  720   &  1.357  &  1.07  \\
And\,XIX    &  2017 09 03  &  8  &  $V$  &  735   &  1.010  &  1.14   &  Cetus       &  2017 10 08  &  9  &  $i$  &  555   &  1.299  &  1.42  \\
And\,XIX    &  2017 10 06  &  9  &  $i$  &  555   &  1.020  &  1.04   &  Palomar\,4  &  2017 01 31  &  6  &  $i$  &  45    &  1.009  &  2.47  \\
And\,XIX    &  2017 10 09  &  9  &  $V$  &  1135  &  1.166  &  1.35   &  Palomar\,4  &  2017 02 01  &  6  &  $V$  &  54    &  1.127  &  1.81  \\

And\,XX     &  2016 02 09  &  2  &  $i$  &  540   &  1.570  &  1.39   &  UGC\,4879   &  2016 02 07  &  2  &  $i$  &  2700  &  1.105  &  1.29  \\
And\,XX     &  2016 06 15  &  3  &  $i$  &  555   &  1.638  &  1.60   &  UGC\,4879   &  2016 02 07  &  2  &  $V$  &  1400  &  1.095  &  1.49  \\
And\,XX     &  2016 08 10  &  4  &  $i$  &  555   &  1.059  &  1.32   &  UGC\,4879   &  2016 02 08  &  2  &  $V$  &  3150  &  1.131  &  1.65  \\
And\,XX     &  2016 08 12  &  4  &  $V$  &  735   &  1.010  &  1.31   &  UGC\,4879   &  2016 06 15  &  3  &  $i$  &  70    &  2.500  &  1.48  \\
And\,XX     &  2016 10 21  &  5  &  $i$  &  540   &  1.037  &  1.09   &  UGC\,4879   &  2016 10 19  &  5  &  $i$  &  2700  &  1.318  &  1.51  \\
And\,XX     &  2016 10 22  &  5  &  $V$  &  735   &  1.025  &  1.12   &  UGC\,4879   &  2016 10 22  &  5  &  $V$  &  1050  &  1.510  &  1.56  \\
And\,XX     &  2017 01 30  &  6  &  $i$  &  555   &  2.302  &  2.99   &  UGC\,4879   &  2018 02 24  &  10 &  $i$  &  2400  &  1.150  &  2.86  \\
And\,XX     &  2017 08 02  &  7  &  $i$  &  555   &  1.225  &  1.11   &  Sextans\,A  &  2016 02 09  &  2  &  $i$  &  2700  &  1.213  &  1.33  \\
And\,XX     &  2017 08 02  &  7  &  $V$  &  735   &  1.155  &  1.34   &  Sextans\,A  &  2016 02 09  &  2  &  $V$  &  3150  &  1.207  &  1.40  \\
And\,XX     &  2017 09 03  &  8  &  $i$  &  555   &  1.033  &  1.04   &  Sextans\,A  &  2017 01 29  &  6  &  $i$  &  1815  &  1.297  &  1.65  \\
And\,XX     &  2017 09 03  &  8  &  $V$  &  735   &  1.061  &  1.08   &  Sextans\,B  &  2016 02 08  &  2  &  $i$  &  2700  &  1.126  &  1.65  \\
And\,XX     &  2017 10 07  &  9  &  $i$  &  555   &  1.018  &  1.44   &  Sextans\,B  &  2016 02 08  &  2  &  $V$  &  2700  &  1.090  &  1.70  \\
And\,XX     &  2017 10 09  &  9  &  $V$  &  735   &  1.441  &  1.39   &  Sextans\,B  &  2017 01 29  &  6  &  $i$  &  2715  &  1.092  &  1.41  \\

And\,XXI    &  2015 06 20  &  1  &  $I$  &  490   &  1.256  &  1.58   &  Sextans\,C  &  2016 02 10  &  2  &  $i$  &  165   &  1.159  &  1.02  \\ 
And\,XXI    &  2015 06 20  &  1  &  $V$  &  540   &  1.187  &  1.61   &  Sextans\,C  &  2016 02 10  &  2  &  $V$  &  165   &  1.172  &  1.15  \\
And\,XXI    &  2016 02 10  &  2  &  $i$  &  630   &  1.911  &  1.66   &  Sextans\,C  &  2017 01 29  &  6  &  $i$  &  45    &  1.186  &  1.42  \\
And\,XXI    &  2016 06 13  &  3  &  $i$  &  615   &  2.296  &  1.70   &  Sextans\,C  &  2017 01 31  &  6  &  $V$  &  45    &  1.165  &  1.88  \\
And\,XXI    &  2016 08 09  &  4  &  $i$  &  555   &  1.088  &  2.46   &  Sextans     &  2016 02 10  &  2  &  $i$  &  90    &  1.217  &  0.94  \\
And\,XXI    &  2016 08 09  &  4  &  $V$  &  735   &  1.060  &  2.66   &  Sextans     &  2016 02 10  &  2  &  $V$  &  180   &  1.239  &  1.25  \\
And\,XXI    &  2016 08 10  &  4  &  $i$  &  555   &  1.099  &  1.37   &  Sextans     &  2017 01 29  &  6  &  $i$  &  45    &  1.340  &  1.88  \\
And\,XXI    &  2016 08 13  &  4  &  $V$  &  735   &  1.452  &  1.75   &  Sextans     &  2017 01 31  &  6  &  $V$  &  45    &  1.165  &  2.80  \\
And\,XXI    &  2016 10 21  &  5  &  $i$  &  555   &  1.070  &  1.40   &  Sextans     &  2018 02 22  &  10 &  $i$  &  45    &  1.165  &  1.94  \\
And\,XXI    &  2016 10 22  &  5  &  $i$  &  540   &  1.031  &  0.98   &  Willman\,1  &  2016 02 07  &  2  &  $i$  &  100   &  1.705  &  1.32  \\
\noalign{\smallskip}
\hline
\end{tabular}
\end{table*}

\begin{table*}
\centering
	%\contcaption{}
\begin{tabular}{lcccccc|lcccccc}
	\hline\hline
 \noalign{\smallskip}
	Galaxy                            &
	Date                              &
	E                                 &
	F                                 &
	$t_{\rm exp}$                     &
        Airmass                           &
        seeing                            &
	Galaxy                            &
	Date                              &
	E                                 &
	F                                 &
	$t_{\rm exp}$                     &
        Airmass                           &
        seeing                            \\
        &
        (y\,m\,d)                &
        &
        &
        (sec)                    &
        &
        (arcsec)
        &
        &               
        (y\,m\,d)                &
        &
        &
        (sec)                    &
        &
        (arcsec)                \\ 
 \noalign{\smallskip}
	\hline
\noalign{\smallskip}
Willman\,1  &  2016 02 07  &  2  &  $V$  &  135   &  1.612  &  1.39   &  Aquarius    &  2016 08 12  &  4  &  $i$  &  558   &  1.360  &  1.38  \\ 
Willman\,1  &  2016 02 08  &  2  &  $i$  &  90    &  1.105  &  1.13   &  Aquarius    &  2016 10 20  &  5  &  $i$  &  2715  &  1.382  &  1.67  \\
Willman\,1  &  2016 02 08  &  2  &  $V$  &  90    &  1.116  &  1.52   &  Aquarius    &  2017 10 06  &  9  &  $i$  &  555   &  1.362  &  1.78  \\ 
Willman\,1  &  2016 06 14  &  3  &  $i$  &  107   &  1.471  &  1.08   &  Aquarius    &  2017 10 08  &  9  &  $V$  &  735   &  1.351  &  1.71  \\
Willman\,1  &  2016 10 21  &  5  &  $i$  &  45    &  1.496  &  1.41   &  CVn\,I      &  2016 02 07  &  2  &  $i$  &  450   &  1.024  &  1.19  \\
Willman\,1  &  2016 10 21  &  5  &  $V$  &  45    &  1.445  &  1.76   &  CVn\,I      &  2016 02 07  &  2  &  $V$  &  675   &  1.014  &  1.48  \\
Willman\,1  &  2017 01 30  &  6  &  $i$  &  45    &  1.192  &  2.46   &  CVn\,I      &  2016 06 14  &  3  &  $i$  &  47    &  1.178  &  1.58  \\
Willman\,1  &  2017 01 31  &  6  &  $V$  &  45    &  1.081  &  2.57   &  CVn\,I      &  2016 08 12  &  4  &  $i$  &  48    &  1.534  &  1.11  \\
Willman\,1  &  2018 02 22  &  10 &  $i$  &  90    &  1.081  &  2.47   &  CVn\,I      &  2016 08 13  &  4  &  $V$  &  45    &  1.541  &  1.29  \\

WLM         &  2016 10 20  &  5  &  $i$  &  1095  &  1.399  &  1.46   &  CVn\,I      &  2017 02 01  &  6  &  $i$  &  45    &  1.227  &  1.48  \\
WLM         &  2016 10 20  &  5  &  $V$  &  1455  &  1.446  &  1.65   &  CVn\,I      &  2017 08 02  &  7  &  $i$  &  45    &  1.439  &  1.17  \\
WLM         &  2016 10 06  &  9  &  $i$  &  1095  &  1.474  &  1.45   &  CVn\,I      &  2017 08 02  &  7  &  $V$  &  45    &  1.381  &  1.44  \\

Pegasus     &  2016 06 14  &  3  &  $i$  &  555   &  1.830  &  2.01   &  CVn\,I      &  2018 02 22  &  10 &  $i$  &  45    &  1.007  &  2.18  \\
Pegasus     &  2016 08 11  &  4  &  $i$  &  555   &  1.256  &  1.49   &  CVn\,I      &  2018 02 22  &  10 &  $V$  &  25    &  1.010  &  2.57  \\
Pegasus     &  2016 08 11  &  4  &  $V$  &  735   &  1.169  &  1.46   &  CVn\,II     &  2015 06 19  &  1  &  $I$  &  660   &  1.087  &  0.82  \\
Pegasus     &  2016 10 20  &  5  &  $i$  &  915   &  1.041  &  1.59   &  CVn\,II     &  2015 06 19  &  1  &  $V$  &  430   &  1.143  &  1.22  \\
Pegasus     &  2016 10 20  &  5  &  $V$  &  1095  &  1.030  &  1.60   &  CVn\,II     &  2016 02 08  &  2  &  $i$  &  90    &  1.005  &  1.73  \\
Pegasus     &  2017 09 03  &  8  &  $i$  &  555   &  1.809  &  1.32   &  CVn\,II     &  2016 02 08  &  2  &  $V$  &  135   &  1.005  &  1.65  \\
Pegasus     &  2017 09 03  &  8  &  $V$  &  735   &  1.568  &  1.50   &  CVn\,II     &  2016 02 10  &  2  &  $i$  &  240   &  1.015  &  1.35  \\
Pegasus     &  2017 10 07  &  9  &  $i$  &  555   &  1.073  &  2.34   &  CVn\,II     &  2016 02 10  &  2  &  $V$  &  360   &  1.022  &  1.40  \\

Pisces\,I   &  2016 08 13  &  4  &  $i$  &  555   &  1.225  &  1.48   &  CVn\,II     &  2016 06 14  &  3  &  $i$  &  47    &  1.188  &  1.40  \\  
Pisces\,I   &  2016 10 20  &  5  &  $i$  &  555   &  1.378  &  1.32   &  CVn\,II     &  2017 02 01  &  6  &  $i$  &  45    &  1.127  &  1.58  \\
Pisces\,I   &  2016 10 20  &  5  &  $V$  &  735   &  1.527  &  1.66   &  CVn\,II     &  2018 02 22  &  10 &  $i$  &  45    &  1.016  &  2.97  \\
Pisces\,I   &  2017 01 29  &  6  &  $i$  &  555   &  1.279  &  1.55   &  CVn\,II     &  2018 02 22  &  10 &  $V$  &  60    &  1.009  &  2.87  \\
Pisces\,I   &  2017 10 07  &  9  &  $i$  &  555   &  1.477  &  1.66   &  Coma\,Ber   &  2015 06 20  &  1  &  $I$  &  100   &  1.114  &  0.81  \\

Pisces\,II  &  2015 06 17  &  1  &  $I$  &  450   &  1.468  &  1.02   &  Coma\,Ber   &  2015 06 20  &  1  &  $V$  &  270   &  1.231  &  1.18  \\
Pisces\,II  &  2015 06 17  &  1  &  $V$  &  630   &  1.341  &  1.29   &  Coma\,Ber   &  2016 02 08  &  2  &  $i$  &  45    &  1.023  &  1.19  \\
Pisces\,II  &  2016 06 15  &  3  &  $i$  &  55    &  1.595  &  1.66   &  Coma\,Ber   &  2016 02 08  &  2  &  $V$  &  45    &  1.033  &  1.52  \\
Pisces\,II  &  2016 08 11  &  4  &  $i$  &  72    &  1.502  &  1.34   &  Coma\,Ber   &  2016 02 10  &  2  &  $i$  &  120   &  1.022  &  1.37  \\
Pisces\,II  &  2016 08 11  &  4  &  $V$  &  72    &  1.361  &  1.52   &  Coma\,Ber   &  2016 02 10  &  2  &  $V$  &  120   &  1.030  &  1.30  \\
Pisces\,II  &  2016 08 13  &  4  &  $i$  &  47    &  1.553  &  1.43   &  Coma\,Ber   &  2016 06 14  &  3  &  $i$  &  38    &  1.266  &  1.42  \\
Pisces\,II  &  2016 10 21  &  5  &  $i$  &  90    &  1.113  &  1.16   &  Coma\,Ber   &  2017 02 01  &  6  &  $i$  &  36    &  1.227  &  1.58  \\
Pisces\,II  &  2016 10 21  &  5  &  $V$  &  135   &  1.101  &  1.55   &  Coma\,Ber   &  2017 02 01  &  6  &  $V$  &  20    &  1.239  &  2.24  \\
Pisces\,II  &  2017 10 07  &  9  &  $i$  &  45    &  1.522  &  1.91   &  Coma\,Ber   &  2018 02 22  &  10 &  $i$  &  36    &  1.007  &  2.64  \\

Segue\,I    &  2016 02 08  &  2  &  $i$  &  45    &  1.199  &  1.32   &  Coma\,Ber   &  2018 02 22  &  10 &  $V$  &  36    &  1.004  &  2.80  \\
Segue\,I    &  2016 02 08  &  2  &  $V$  &  45    &  1.232  &  1.70   &  Leo\,I      &  2016 02 09  &  2  &  $i$  &  135   &  1.109  &  1.01  \\
Segue\,I    &  2016 02 09  &  2  &  $i$  &  45    &  1.197  &  1.16   &  Leo\,I      &  2016 02 09  &  2  &  $V$  &  180   &  1.089  &  1.15  \\
Segue\,I    &  2016 02 09  &  2  &  $V$  &  45    &  1.158  &  1.23   &  Leo\,I      &  2016 02 10  &  2  &  $i$  &  180   &  1.245  &  1.37  \\
Segue\,I    &  2016 06 14  &  3  &  $i$  &  40    &  1.759  &  1.22   &  Leo\,I      &  2016 02 10  &  2  &  $V$  &  240   &  1.280  &  1.47  \\
Segue\,I    &  2016 10 20  &  5  &  $i$  &  45    &  1.360  &  1.66   &  Leo\,I      &  2016 06 15  &  3  &  $i$  &  25    &  2.825  &  1.31  \\
Segue\,I    &  2017 01 29  &  6  &  $i$  &  36    &  1.203  &  1.78   &  Leo\,I      &  2016 10 19  &  5  &  $i$  &  135   &  1.381  &  1.66  \\
Segue\,I    &  2017 01 31  &  6  &  $V$  &  36    &  1.030  &  1.72   &  Leo\,I      &  2017 01 29  &  6  &  $i$  &  35    &  1.317  &  1.98  \\
Segue\,I    &  2018 02 22  &  10 &  $i$  &  36    &  1.026  &  2.01   &  Leo\,I      &  2017 01 31  &  6  &  $V$  &  73    &  1.044  &  1.57  \\

Segue\,II   &  2016 02 08  &  2  &  $i$  &  45    &  1.192  &  1.78   &  Leo\,II     &  2016 02 07  &  2  &  $i$  &  90    &  1.022  &  1.01  \\
Segue\,II   &  2016 10 19  &  5  &  $i$  &  45    &  1.356  &  0.91   &  Leo\,II     &  2016 02 07  &  2  &  $V$  &  135   &  1.039  &  1.22  \\
Segue\,II   &  2016 10 19  &  5  &  $V$  &  49    &  1.430  &  1.05   &  Leo\,II     &  2016 02 08  &  2  &  $i$  &  90    &  1.209  &  1.07  \\
Segue\,II   &  2017 01 29  &  6  &  $i$  &  50    &  1.056  &  1.73   &  Leo\,II     &  2016 02 08  &  2  &  $V$  &  135   &  1.250  &  1.01  \\ 
Segue\,II   &  2017 09 01  &  8  &  $i$  &  36    &  1.042  &  1.19   &  Leo\,II     &  2016 02 10  &  2  &  $i$  &  240   &  1.262  &  1.00  \\
Segue\,II   &  2017 09 01  &  8  &  $V$  &  45    &  1.056  &  1.35   &  Leo\,II     &  2016 02 10  &  2  &  $V$  &  360   &  1.307  &  1.37  \\
Segue\,II   &  2017 10 07  &  9  &  $i$  &  36    &  1.251  &  1.55   &  Leo\,II     &  2016 06 14  &  3  &  $i$  &  47    &  1.610  &  1.44  \\
 
Segue\,III  &  2015 06 17  &  1  &  $I$  &  45    &  1.209  &  0.99   &  Leo\,II     &  2017 01 29  &  6  &  $i$  &  68    &  1.135  &  1.55  \\ 
Segue\,III  &  2015 06 17  &  1  &  $V$  &  450   &  1.264  &  1.25   &  Leo\,II     &  2017 02 01  &  6  &  $V$  &  63    &  1.227  &  2.08  \\
Segue\,III  &  2016 06 13  &  3  &  $i$  &  30    &  2.330  &  1.62   &  Leo\,II     &  2018 02 22  &  10 &  $i$  &  15    &  1.008  &  2.54  \\
Segue\,III  &  2016 06 13  &  3  &  $V$  &  27    &  2.132  &  1.98   &  Leo\,A      &  2016 02 09  &  2  &  $i$  &  540   &  1.252  &  1.56  \\
Segue\,III  &  2016 08 10  &  4  &  $i$  &  27    &  1.031  &  0.82   &  Leo\,A      &  2016 02 09  &  2  &  $V$  &  720   &  1.180  &  1.69  \\
Segue\,III  &  2016 08 10  &  4  &  $V$  &  27    &  1.021  &  1.42   &  Leo\,A      &  2016 02 10  &  2  &  $i$  &  600   &  1.192  &  1.68  \\
Segue\,III  &  2016 08 13  &  4  &  $i$  &  27    &  1.171  &  1.58   &  Leo\,A      &  2016 02 10  &  2  &  $V$  &  900   &  1.143  &  1.67  \\
Segue\,III  &  2016 10 21  &  5  &  $i$  &  42    &  1.016  &  1.32   &  Leo\,A      &  2016 06 15  &  3  &  $i$  &  555   &  2.050  &  1.20  \\
Segue\,III  &  2016 10 21  &  5  &  $V$  &  45    &  1.021  &  1.22   &  Leo\,A      &  2016 10 20  &  5  &  $i$  &  555   &  1.448  &  1.87  \\
Segue\,III  &  2017 08 02  &  7  &  $i$  &  27    &  1.070  &  1.16   &  Leo\,A      &  2016 10 20  &  5  &  $V$  &  735   &  1.322  &  1.84  \\
Segue\,III  &  2017 08 02  &  7  &  $V$  &  27    &  1.055  &  1.09   &  Leo\,A      &  2017 01 29  &  6  &  $i$  &  555   &  1.049  &  1.56  \\
Segue\,III  &  2017 10 07  &  9  &  $i$  &  27    &  1.025  &  2.87   &  Leo\,A      &  2017 01 31  &  6  &  $V$  &  735   &  1.041  &  1.86  \\

SAG\,dig    &  2016 06 13  &  3  &  $i$  &  719   &  1.711  &  2.61   &  Leo\,A      &  2018 02 22  &  10 &  $i$  &  555   &  1.019  &  1.86  \\  
SAG\,dig    &  2016 08 09  &  4  &  $i$  &  555   &  1.449  &  2.22   &  Leo\,T      &  2016 02 07  &  2  &  $i$  &  270   &  1.141  &  1.23  \\
SAG\,dig    &  2016 08 09  &  4  &  $V$  &  735   &  1.460  &  2.38   &  Leo\,T      &  2016 02 07  &  2  &  $V$  &  360   &  1.109  &  1.32  \\
SAG\,dig    &  2016 08 10  &  4  &  $i$  &  555   &  1.564  &  1.23   &  Leo\,T      &  2016 02 10  &  2  &  $i$  &  360   &  1.092  &  1.60  \\
SAG\,dig    &  2016 08 10  &  4  &  $V$  &  735   &  1.471  &  1.64   &  Leo\,T      &  2016 02 10  &  2  &  $V$  &  480   &  1.072  &  1.85  \\
SAG\,dig    &  2016 08 12  &  4  &  $i$  &  558   &  1.449  &  1.37   &  Leo\,T      &  2016 10 19  &  5  &  $i$  &  285   &  1.267  &  1.50  \\
SAG\,dig    &  2016 10 21  &  5  &  $i$  &  2715  &  1.595  &  1.84   &  Leo\,T      &  2016 10 21  &  5  &  $V$  &  360   &  1.455  &  1.77  \\
SAG\,dig    &  2017 08 02  &  7  &  $i$  &  555   &  1.453  &  1.44   &  Leo\,T      &  2017 01 29  &  6  &  $i$  &  365   &  1.080  &  1.71  \\
SAG\,dig    &  2017 08 02  &  7  &  $V$  &  735   &  1.450  &  1.37   &  Leo\,T      &  2017 01 31  &  6  &  $V$  &  375   &  1.071  &  1.90  \\
SAG\,dig    &  2017 09 02  &  8  &  $i$  &  555   &  1.457  &  1.08   &  Leo\,T      &  2018 02 22  &  10 &  $i$  &  285   &  1.022  &  1.99  \\
SAG\,dig    &  2017 09 02  &  8  &  $V$  &  735   &  1.449  &  1.18   &  Leo\,V      &  2016 02 09  &  2  &  $i$  &  90    &  1.150  &  1.13  \\
SAG\,dig    &  2017 10 06  &  9  &  $i$  &  555   &  1.511  &  1.61   &  Leo\,V      &  2016 02 09  &  2  &  $V$  &  135   &  1.166  &  1.41  \\
SAG\,dig    &  2017 10 08  &  9  &  $V$  &  735   &  1.542  &  1.82   &  Leo\,V      &  2016 02 10  &  2  &  $i$  &  140   &  1.169  &  1.33  \\

Aquarius    &  2016 08 10  &  4  &  $i$  &  555   &  1.618  &  1.22   &  Leo\,V      &  2016 02 10  &  2  &  $V$  &  180   &  1.155  &  1.50  \\
Aquarius    &  2016 08 11  &  4  &  $V$  &  735   &  1.649  &  1.62   &  Leo\,V      &  2016 06 14  &  3  &  $i$  &  45    &  1.299  &  1.04  \\
\noalign{\smallskip}
\hline
\end{tabular}
\end{table*}

\begin{table*}
\centering
	%\contcaption{}
\begin{tabular}{lcccccc|lcccccc}
	\hline\hline
 \noalign{\smallskip}
	Galaxy                            &
	Date                              &
	E                                 &
	F                                 &
	$t_{\rm exp}$                     &
        Airmass                           &
        seeing                            &
	Galaxy                            &
	Date                              &
	E                                 &
	F                                 &
	$t_{\rm exp}$                     &
        Airmass                           &
        seeing                            \\
        &
        (y\,m\,d)                &
        &
        &
        (sec)                    &
        &
        (arcsec)
        &
        &               
        (y\,m\,d)                &
        &
        &
        (sec)                    &
        &
        (arcsec)                \\ 
 \noalign{\smallskip}
	\hline
\noalign{\smallskip}
Leo\,V      &  2017 01 30  &  6  &  $i$  &  45    &  1.379  &  2.14   &  Bootes\,II  &  2017 08 01  &  7  &  $V$  &  45    &  1.718  &  1.66  \\
Leo\,V      &  2017 02 01  &  6  &  $V$  &  63    &  1.327  &  2.08   &  Bootes\,III &  2016 02 09  &  2  &  $i$  &  45    &  1.001  &  0.96  \\
Leo\,V      &  2018 02 22  &  10 &  $i$  &  45    &  1.124  &  1.88   &  Bootes\,III &  2016 02 09  &  2  &  $V$  &  45    &  1.003  &  1.15  \\

Leo\,IV     &  2016 02 09  &  2  &  $i$  &  90    &  1.217  &  1.08   &  Bootes\,III &  2016 06 13  &  3  &  $i$  &  58    &  1.007  &  1.12  \\
Leo\,IV     &  2016 02 09  &  2  &  $V$  &  135   &  1.241  &  1.06   &  Bootes\,III &  2016 06 13  &  3  &  $V$  &  42    &  1.013  &  1.13  \\
Leo\,IV     &  2016 02 10  &  2  &  $i$  &  120   &  1.220  &  1.13   &  Bootes\,III &  2016 08 12  &  4  &  $i$  &  39    &  1.580  &  1.12  \\
Leo\,IV     &  2016 02 10  &  2  &  $V$  &  180   &  1.239  &  1.75   &  Bootes\,III &  2016 08 13  &  4  &  $V$  &  45    &  1.660  &  1.31  \\
Leo\,IV     &  2016 06 14  &  3  &  $i$  &  45    &  1.382  &  1.02   &  Bootes\,III &  2017 01 29  &  6  &  $i$  &  36    &  1.002  &  2.31  \\
Leo\,IV     &  2017 01 30  &  6  &  $i$  &  45    &  1.375  &  3.33   &  Bootes\,III &  2017 08 01  &  7  &  $i$  &  36    &  1.633  &  1.55  \\
Leo\,IV     &  2017 02 01  &  6  &  $V$  &  63    &  1.325  &  1.62   &  Bootes\,III &  2017 08 01  &  7  &  $V$  &  36    &  1.732  &  1.73  \\
Leo\,IV     &  2018 02 22  &  10 &  $i$  &  45    &  1.148  &  1.91   &  Hercules    &  2015 06 18  &  1  &  $I$  &  90    &  1.165  &  0.89  \\

Leo\,P      &  2016 02 10  &  2  &  $i$  &  2400  &  1.066  &  1.46   &  Hercules    &  2015 06 18  &  1  &  $V$  &  450   &  1.274  &  0.82  \\ 
Leo\,P      &  2016 02 10  &  2  &  $V$  &  2250  &  1.027  &  1.57   &  Hercules    &  2016 02 09  &  2  &  $i$  &  63    &  1.187  &  1.12  \\

UMa\,I      &  2016 02 09  &  2  &  $i$  &  45    &  1.206  &  1.12   &  Hercules    &  2016 02 09  &  2  &  $V$  &  90    &  1.159  &  1.25  \\
UMa\,I      &  2016 02 09  &  2  &  $V$  &  45    &  1.186  &  1.37   &  Hercules    &  2016 06 13  &  3  &  $i$  &  50    &  1.088  &  1.13  \\
UMa\,I      &  2016 02 10  &  2  &  $i$  &  120   &  1.144  &  1.06   &  Hercules    &  2016 06 14  &  3  &  $i$  &  47    &  1.042  &  1.35  \\
UMa\,I      &  2016 02 10  &  2  &  $V$  &  120   &  1.155  &  1.09   &  Hercules    &  2016 08 09  &  4  &  $i$  &  45    &  1.059  &  1.44  \\
UMa\,I      &  2016 06 14  &  3  &  $i$  &  49    &  1.465  &  1.01   &  Hercules    &  2016 08 09  &  4  &  $V$  &  45    &  1.070  &  2.66  \\
UMa\,I      &  2016 10 21  &  5  &  $i$  &  45    &  1.544  &  1.41   &  Hercules    &  2016 08 10  &  4  &  $i$  &  45    &  1.059  &  0.99  \\
UMa\,I      &  2016 10 21  &  5  &  $V$  &  45    &  1.490  &  1.97   &  Hercules    &  2016 08 11  &  4  &  $i$  &  27    &  1.103  &  1.02  \\
UMa\,I      &  2017 01 29  &  6  &  $i$  &  45    &  1.227  &  1.42   &  Hercules    &  2016 08 11  &  4  &  $V$  &  51    &  1.126  &  1.15  \\
UMa\,I      &  2017 01 31  &  6  &  $V$  &  45    &  1.090  &  2.05   &  Hercules    &  2016 08 12  &  4  &  $i$  &  48    &  1.304  &  1.02  \\
UMa\,I      &  2018 02 22  &  10 &  $i$  &  45    &  1.087  &  2.28   &  Hercules    &  2016 10 22  &  5  &  $i$  &  63    &  1.873  &  1.68  \\

UMa\,II     &  2016 02 07  &  2  &  $i$  &  45    &  1.431  &  1.19   &  Hercules    &  2017 02 01  &  6  &  $i$  &  45    &  1.127  &  1.29  \\ 
UMa\,II     &  2016 02 07  &  2  &  $V$  &  45    &  1.404  &  1.52   &  Hercules    &  2017 08 02  &  7  &  $i$  &  45    &  1.152  &  1.21  \\
UMa\,II     &  2016 02 10  &  2  &  $i$  &  120   &  1.277  &  1.40   &  Hercules    &  2017 08 02  &  7  &  $V$  &  45    &  1.127  &  1.46  \\ 
UMa\,II     &  2016 02 10  &  2  &  $V$  &  120   &  1.267  &  1.43   &  Hercules    &  2017 09 03  &  8  &  $i$  &  555   &  1.198  &  0.96  \\
UMa\,II     &  2016 10 20  &  5  &  $i$  &  45    &  1.406  &  1.35   &  Hercules    &  2017 09 03  &  8  &  $V$  &  735   &  1.281  &  1.00  \\
UMa\,II     &  2016 10 20  &  5  &  $V$  &  45    &  1.379  &  1.70   &  Ursa\,minor &  2015 06 20  &  1  &  $I$  &  130   &  1.284  &  1.02  \\
UMa\,II     &  2017 01 29  &  6  &  $i$  &  36    &  1.230  &  1.17   &  Ursa\,minor &  2015 06 20  &  1  &  $V$  &  450   &  1.309  &  1.12  \\
UMa\,II     &  2017 01 31  &  6  &  $V$  &  36    &  1.225  &  1.83   &  Ursa\,minor &  2016 02 10  &  2  &  $i$  &  165   &  1.286  &  1.41  \\
UMa\,II     &  2017 10 06  &  9  &  $i$  &  36    &  1.490  &  1.40   &  Ursa\,minor &  2016 02 10  &  2  &  $V$  &  165   &  1.282  &  1.55  \\
UMa\,II     &  2017 10 09  &  9  &  $V$  &  36    &  1.394  &  1.27   &  Ursa\,minor &  2016 06 13  &  3  &  $i$  &  40    &  1.276  &  0.86  \\
UMa\,II     &  2018 02 22  &  10 &  $i$  &  36    &  1.210  &  1.81   &  Ursa\,minor &  2016 08 12  &  4  &  $i$  &  78    &  1.447  &  1.12  \\
UMa\,II     &  2018 02 24  &  10 &  $V$  &  36    &  1.272  &  3.99   &  Ursa\,minor &  2016 08 12  &  4  &  $V$  &  36    &  1.474  &  1.73  \\

Bootes\,I   &  2015 06 18  &  1  &  $I$  &  50    &  1.032  &  0.92   &  Ursa\,minor &  2016 10 22  &  5  &  $i$  &  45    &  2.083  &  1.48  \\ 
Bootes\,I   &  2015 06 18  &  1  &  $V$  &  540   &  1.046  &  1.55   &  Ursa\,minor &  2017 02 01  &  6  &  $i$  &  36    &  1.227  &  0.92  \\
Bootes\,I   &  2016 02 09  &  2  &  $i$  &  45    &  1.049  &  1.17   &  Ursa\,minor &  2017 08 02  &  7  &  $i$  &  36    &  1.385  &  1.09  \\
Bootes\,I   &  2016 02 09  &  2  &  $V$  &  45    &  1.041  &  1.25   &  Ursa\,minor &  2017 08 02  &  7  &  $V$  &  36    &  1.366  &  1.29  \\
Bootes\,I   &  2016 06 14  &  3  &  $i$  &  38    &  1.130  &  1.45   &  Draco       &  2015 06 19  &  1  &  $I$  &  210   &  1.192  &  0.92  \\
Bootes\,I   &  2016 08 10  &  4  &  $i$  &  45    &  1.572  &  1.14   &  Draco       &  2015 06 19  &  1  &  $V$  &  420   &  1.247  &  1.15  \\
Bootes\,I   &  2016 08 13  &  4  &  $V$  &  36    &  1.601  &  1.23   &  Draco       &  2016 06 13  &  3  &  $i$  &  56    &  1.193  &  1.73  \\
Bootes\,I   &  2017 02 01  &  6  &  $i$  &  36    &  1.227  &  1.72   &  Draco       &  2016 06 13  &  3  &  $V$  &  45    &  1.178  &  2.38  \\
Bootes\,I   &  2017 08 01  &  7  &  $i$  &  36    &  1.422  &  1.42   &  Draco       &  2016 08 09  &  4  &  $i$  &  135   &  1.150  &  2.51  \\ 
Bootes\,I   &  2017 08 01  &  7  &  $V$  &  36    &  1.492  &  1.67   &  Draco       &  2016 08 09  &  4  &  $V$  &  135   &  1.153  &  2.84  \\

Bootes\,II  &  2015 06 19  &  1  &  $I$  &  95    &  1.215  &  0.76   &  Draco       &  2016 08 10  &  4  &  $i$  &  45    &  1.149  &  1.22  \\
Bootes\,II  &  2015 06 19  &  1  &  $V$  &  380   &  1.327  &  1.32   &  Draco       &  2016 08 11  &  4  &  $i$  &  27    &  1.145  &  0.86  \\
Bootes\,II  &  2016 02 09  &  2  &  $i$  &  54    &  1.043  &  1.06   &  Draco       &  2016 08 11  &  4  &  $V$  &  51    &  1.146  &  1.13  \\
Bootes\,II  &  2016 02 09  &  2  &  $V$  &  72    &  1.041  &  1.19   &  Draco       &  2016 08 12  &  4  &  $i$  &  48    &  1.173  &  1.06  \\
Bootes\,II  &  2016 06 15  &  3  &  $i$  &  30    &  1.122  &  1.25   &  Draco       &  2016 10 20  &  5  &  $i$  &  45    &  1.485  &  1.63  \\
Bootes\,II  &  2016 08 12  &  4  &  $i$  &  48    &  1.647  &  1.09   &  Draco       &  2017 02 01  &  6  &  $i$  &  45    &  1.227  &  1.02  \\
Bootes\,II  &  2016 08 13  &  4  &  $V$  &  45    &  1.752  &  1.24   &  Draco       &  2017 08 02  &  7  &  $i$  &  45    &  1.151  &  1.07  \\
Bootes\,II  &  2017 01 29  &  6  &  $i$  &  45    &  1.047  &  3.30   &  Draco       &  2017 08 02  &  7  &  $V$  &  45    &  1.148  &  0.93  \\
Bootes\,II  &  2017 08 01  &  7  &  $i$  &  45    &  1.616  &  1.49   &  Draco       &  2017 10 07  &  9  &  $i$  &  45    &  1.331  &  1.63  \\
\noalign{\smallskip}
\hline
\end{tabular}
\end{table*}

\label{lastpage}
\end{document}